\documentclass[reprint,superscriptaddress,twocolumn,10pt]{revtex4-1}
\usepackage{bbm}
\usepackage{physics}
\usepackage{amsmath}
\usepackage{epsfig}
\usepackage{subfigure,mathrsfs}
\usepackage{array}
\usepackage{amssymb}
\usepackage{braket}
\usepackage{float}
\usepackage{lmodern,amssymb}
\usepackage{physics}
\usepackage[dvipsnames]{xcolor}
\usepackage{lipsum}

\newcommand{\beq}[0]{\begin{equation}}
\newcommand{\eeq}[0]{\end{equation}}
\newcommand{\non}{\nonumber}
\def\be{\begin{equation}}
\def\ee{\end{equation}}
\def\bea{\begin{eqnarray}}
\def\eea{\end{eqnarray}}
\newcommand{\ba}{\begin{eqnarray}}
\newcommand{\ea}{\end{eqnarray}}
\usepackage{hyperref}

\begin{document}

\title{Floquet quantum thermal transistor}

\author{Nikhil Gupt}
\affiliation{Indian Institute of Technology Kanpur, 
Kanpur,Uttar Pradesh 208016, India}
\author{Srijan Bhattacharyya}
\affiliation{Indian Institute of Technology Kanpur, 
Kanpur,Uttar Pradesh 208016, India}
\affiliation{Department of Chemistry, University of Colorado, Boulder, 
Colorado 80309, USA}
\author{{Bikash Das}}
\affiliation{School of Physical Sciences, Indian Association for the Cultivation of Science, 2A and B Raja S. C. Mullick Road, Jadavpur, Kolkata 700032, India}
\author{{Subhadeep Datta}}
\affiliation{School of Physical Sciences, Indian Association for the Cultivation of Science, 2A and B Raja S. C. Mullick Road, Jadavpur, Kolkata 700032, India}
\author{{Victor Mukherjee}}
\thanks{mukherjeev@iiserbpr.ac.in}
\affiliation{Department of Physical Sciences, IISER Berhampur, Berhampur 760010, India}
\author{{Arnab Ghosh}}
\thanks{arnab@iitk.ac.in}
\affiliation{Indian Institute of Technology Kanpur, 
Kanpur,Uttar Pradesh 208016, India}
\begin{abstract}
We apply periodic control to realize a quantum thermal transistor, which we term as the Floquet Quantum thermal Transistor. Periodic modulation allows us to control the heat flows and achieve large amplification factors even for fixed bath temperatures. Importantly, this transistor effect persists in the {\it cut-off region}, where traditional quantum thermal transistors operating in absence of periodic modulation, fail to act as viable heat modulation devices.
\end{abstract}

\maketitle

\section{INTRODUCTION}

Research on quantum machines \cite{millen2016rise, mukherjee21many, myers2022quantum, arnab2019arequantum, sebastian2019book} has received immense interest during the last decade, owing to the important role played by them in the fields of quantum thermodynamics \cite{gelbwaser2015thermodynamics, arnab2018two, vinjanampathy2016quantum, binder2019thermodynamics, bhattacharjee21quantum, nikhil2021statistical} and quantum technologies \cite{kurizki2015quantum, arnab2017catalysis}. Recent progress in the ability of researchers to probe and control systems in the quantum regime \cite{rosi13fast, borselli21two} have led to experimental realizations of several such machines \cite{rossnagel2016single, klaers2017squeezed, klatzow19experimental, maslennikov19quantum}. Development of high-performing quantum thermal machines demands improvement in our abilities to control thermal currents at the microscopic scale. Consequently, several studies have focussed on modelling of quantum thermal rectifiers and quantum thermal transistors, aimed at controlling thermal currents in open quantum systems \cite{lashkaryov1941investigations, jiang2015phonon, joulain2016quantum, majland2020quantum, mandarino2021thermal, wijesekara2021darlington}.
Above mentioned thermal rectifiers are analogous to their electronic counterparts; they exhibit asymmetric fluxes when the temperatures at the two ends are inverted. A thermal transistor on the other hand regulates the heat flow between two of its terminals in response to the temperature change of a thermal bath coupled to a third terminal. Following the theoretical proposal of a thermal transistor  reported in~\cite{chung2008thermal, li2006negative}, several works have addressed other types of thermal transistors, including the metal-superconductor thermal transistors~\cite{saira2007heat}, near-and-far-field thermal transistors~\cite{prod'homme2016optimized,
ordonez2016transistorlike,
joulain2015modulation,ben2014near}, and quantum thermal transistors~\cite{joulain2016quantum} to name  a few. A significant advantage of such thermal transistors is that they can implement several thermal operations, viz. heat swap~\cite{bosisio2015magnetic}, heat path selector ~\cite{zhang2017three} etc.

Recent works have shown the possibility to realizing thermal transistor-effect through simple quantum systems. For example, Joulain et al.~\cite{joulain2016quantum} and Mandarino et al.~\cite{mandarino2021thermal}
demonstrated transistor action with three strongly coupled two-level systems (TLSs) interacting with their respective thermal baths. Guo et al.~\cite{guo2019multifunctional} identified similar transistor behaviour with three-qubit system under tri-linear qubit interactions. Zhang et al.~\cite{zhang2018coulomb,zhang2017three} demonstrated that a system of three Coulomb coupled quantum dots can also exhibits such characteristics. In Ref. ~\cite{guo2019multifunctional}, a coupled qubit and qutrit has been shown to display transistor-action in the thermal domain, while Majland et al.~\cite{majland2020quantum} extended it further with superconducting circuits. With an external optical field, the concept of field effect quantum transistor and Darlington transistor have been proposed by Wijesekara et al.~\cite{wijesekara2021darlington}.

Most of the works on quantum thermal transistors till now have focussed on realizing the transistor effect through changes in bath temperatures~\cite{joulain2016quantum, ben2014near, li2006negative, guo2018quantum, zhang2018coulomb, ghosh2021quantum, mandarino2021thermal, naseem2020minimal}. In contrast, in this paper we apply quantum control to realize a model of a \emph{Floquet Quantum thermal Transistor} (FQT); here we achieve the transistor effect through periodic modulation of the system Hamiltonian. This periodic modulation allows us to control the heat currents even for fixed bath temperatures. Most importantly, the present control scheme enables one to achieve the transistor effect even in the so called {\it cut-off} regime, where traditional quantum thermal transistors, which depend on variation of bath temperatures, fail to perform.

\par The present work is organized as follows: We introduce the model, dynamics  and thermal currents in Sec. \ref{secII},  Sec. \ref{Sec.IIIA} deals with general transistor characteristics, we give a comparison with an unmodulated quantum thermal transistor in Sec. \ref{Sec.IIIB}, consider FQT at arbitrary temperatures in Sec. \ref{Sec.IIIC}, and the low-temperature limit in Sec. \ref{Sec.IIID}. Finally, we conclude in Sec. \ref{secIV}.
We include technical details in the Appendix.
\section{Model and Dynamics}
\label{secII}

\begin{figure} 
\centering
\includegraphics[width=\columnwidth]{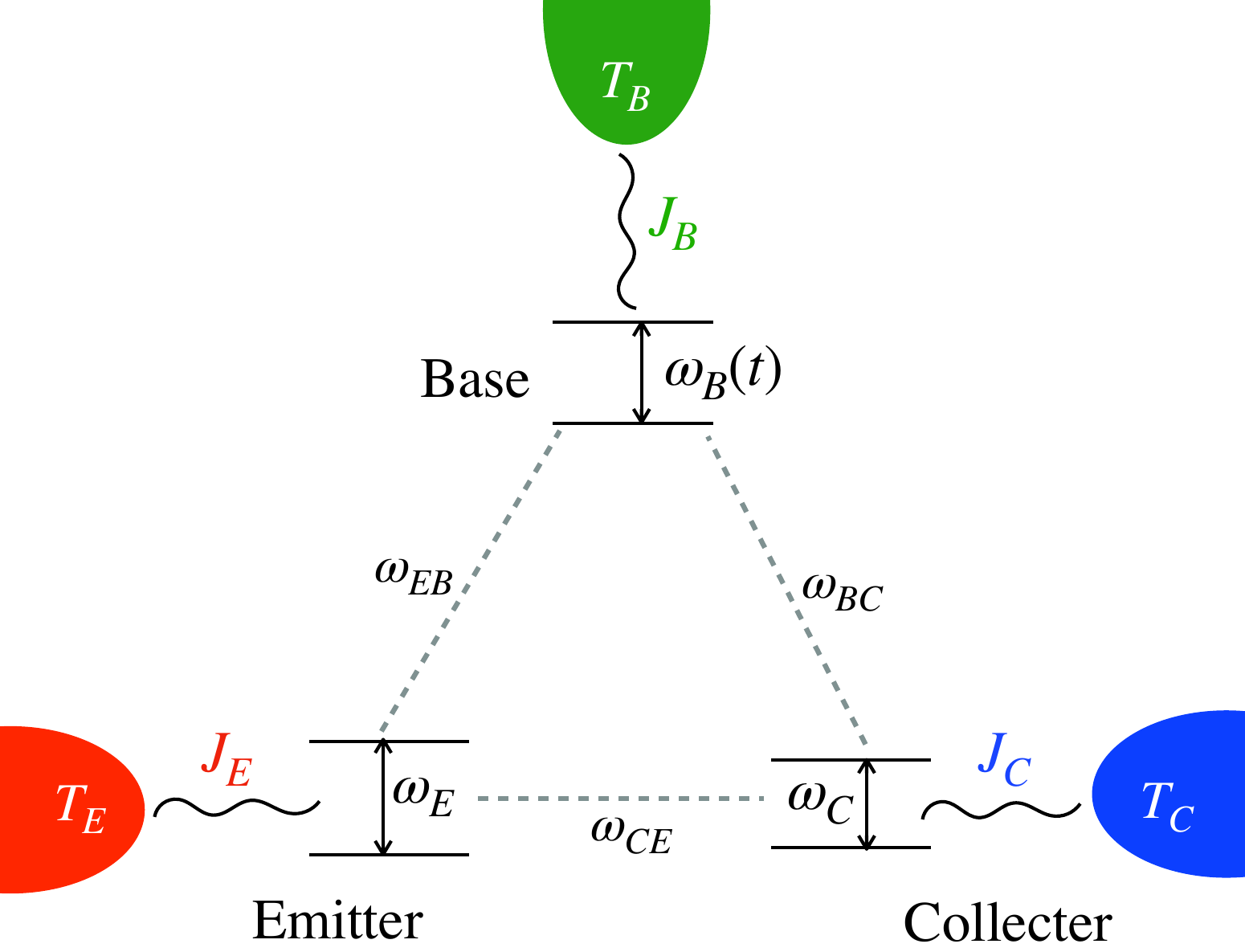}
\caption{ Schematic diagram of the FQT model system. The three terminals (qubits) are coupled to three thermal baths. We achieve thermal transistor effect through periodic modulation of the frequency $\omega_{ B}(t)$ of the base qubit.} 
\label{FQT_fig1}
\end{figure}

Quantum thermal transistors are analogous to their electronic counterparts, with temperatures replacing voltages and thermal
energy flows replacing electric currents \cite{joulain2016quantum}. Quantum thermal transistors are typically three-terminal devices, with the channels coupled to their respective thermal baths (see Fig.~\ref{FQT_fig1}). Heat current enters the setup  through one terminal (termed as the {\it emitter} ($E$))  and leaves through the remaining two terminals (termed as the {\it base} ($B$) and the {\it collector} ($C$)).
Transistor effect is said to be obtained when a small thermal current at the base can be used to control large collector and emitter currents, which can also be modulated, switched and amplified by small changes in the base current. Previous works have studied the modulation of $J_B$ through: (i)  the variation of base temperature as in the case of quantum thermal transistors~\cite{joulain2016quantum} or (ii) by application of external optical field as in the case of field effect quantum transistors~\cite{wijesekara2021darlington}. In contrast, here we shall do the same through periodic modulation of the base terminal frequency. Remarkably, as we discuss below, this periodic modulation allows us to achieve transistor effect even in regimes hitherto considered to be unfavorable for transistor operation.  We note that here we use the sign convention that thermal currents which enter (leave)  the system are positive (negative).

\par We model the Floquet Quantum thermal Transistor (FQT) through three interacting two-level systems (TLSs), representing the three terminals. As shown in Fig.~\ref{FQT_fig1}, the $\alpha$-th TLS is coupled to a thermal reservoir at temperature $T_{ \alpha}$ ($\alpha = {E, B, C}$). Similar to Ref.~\cite{joulain2016quantum}, we consider $T_{ E}$ and $T_{C}$ to be fixed;  $T_{B}$ can vary, and $T_{ E} > T_{B}, T_{ C}$. The Hamiltonian of the whole setup is then given by three terminals FQT system as
\begin{eqnarray}
\label{model-Hamiltonian}
H(t) &=& H_{S}(t) + H_{R} + H_{I}\non\\
H_S(t)&=&\frac{\hbar\omega_E}{2}\sigma_z^E+\frac{\hbar\omega_B(t)}{2}\sigma_z^B+\frac{\hbar\omega_C}{2}\sigma_z^C +\frac{\hbar\omega_{EB}}{2}\sigma_z^E\sigma_z^{B} \non\\
&+&\frac{\hbar\omega_{BC}}{2}\sigma_z^B\sigma_z^{C}+\frac{\hbar\omega_{CE}}{2}\sigma_z^C\sigma_z^{E},\non\\
H_{R} &=& H_{RC} + H_{RB} + H_{RE}, \non\\
H_{I} &=& \sigma_{x}^{C}\otimes \mathcal{R}_{C} + \sigma_{x}^{B}\otimes \mathcal{R}_{B} + \sigma_{x}^{E}\otimes \mathcal{R}_{E}.
\end{eqnarray}
Here $H_{S}(t)$ denotes the Hamiltonian describing the three TLSs; $\sigma_{\theta}^{\alpha}$ is the Pauli matrix along dimension $\theta~(= x, y, z)$ acting on the terminal $\alpha~( = E,B,C)$,  $\omega_\alpha$ refers to the frequency of the respective TLS, whereas $\omega_{\alpha \alpha^{\prime}}$ denotes the interaction strength between the $\alpha$ and $\alpha^{\prime}$ terminals. We assume that the base terminal is periodically modulated at a frequency $\nu=\frac{2\pi}{\tau}$, such that
\begin{equation}\label{omega_B(t)}
\omega_B(t+\tau)= \omega_B(t);\quad  \frac{1}{\tau} \int_t^{t+\tau}\omega_B(t) dt= \omega_0. 
\end{equation}
Here $\omega_0$ denotes the time-averaged frequency of $B$.

The eigenstates of $H_S(t)$, labelled by $\left|1\right>=\left|\uparrow\uparrow\uparrow\right>$, $\left|2\right>=\left|\uparrow\uparrow\downarrow\right>$, $\left|3\right>=\left|\uparrow\downarrow\uparrow\right>$, $\left|4\right>=\left|\uparrow\downarrow\downarrow\right>$, $\left|5\right>=\left|\downarrow\uparrow\uparrow\right>$, $\left|6\right>=\left|\downarrow\uparrow\downarrow\right>$, $\left|7\right>=\left|\downarrow\downarrow\uparrow\right>$ and $\left|8\right>=\left|\downarrow\downarrow\downarrow\right>$~\cite{joulain2016quantum}, are obtained through the tensor product of the $\sigma_z^\alpha$ eigenstates, $\left|\uparrow \right>$ and $\left|\downarrow \right>$ of the individual TLS (see App. \ref{Appendix-A}). $H_{R\alpha}$ denotes the Hamiltonian of the thermal bath coupled to the terminal $\alpha$, through the term  $\sigma_{x}^{\alpha}\otimes \mathcal{R}_{\alpha}$ included in the interaction Hamiltonian  $H_{I}$; $\mathcal{R}_{\alpha}$ is an operator acting on the bath  $\alpha$.

The above interaction imposes restriction on flipping more than one spin at a time. Consequently, there are in total twelve allowed transitions. The base reservoir induces the four transitions between $1\leftrightarrow3$, $2\leftrightarrow 4$, $5\leftrightarrow7$, and $6\leftrightarrow8$, while the emitter bath drives the transitions $1\leftrightarrow5$, $2\leftrightarrow 6$, $3\leftrightarrow7$, and $4\leftrightarrow8$, and the collector bath triggers the transitions $1\leftrightarrow2$, $3\leftrightarrow 4$, $5\leftrightarrow6$, and $7\leftrightarrow8$. The rates at which the above transitions occur can be computed using Floquet-Lindblad master equation~\cite{gelbwaser2015thermodynamics,binder2019thermodynamics,floquet1883equations, mondal2020exploring}.

One can use the Floquet method and implement the Born, Markov and secular approximations, to arrive at the time-independent Lindblad dissipators describing the time-evolution of the density matrix $\rho$ for the state of the FQT, given by
\begin{equation}\label{master_eq}
\frac{d \rho}{dt}=\mathcal{L}_E[\rho]+
\tilde{ \mathcal{L}}_B[\rho]+\mathcal{L}_C[\rho].
\end{equation}
Here the Lindbladians are given by (see App. \ref{Appendix-A}) 
\begin{eqnarray}\label{lindblad}
\mathcal{L}_{\alpha}[\rho] &=& \sum_{\{\Omega_\alpha\}}\big[G_{\alpha}(\Omega_\alpha)\mathcal{D}(A_{\alpha})[\rho] \non\\ 
&+& G_{\alpha}(-\Omega_\alpha)\mathcal{D}(A_{\alpha}^{\dagger})[\rho]\big],\;~~~~ \alpha \in \{E,C\}\nonumber\\
\tilde{\mathcal{L}}_B[\rho]&=&
\sum_{\{\Omega_B\}}\sum_{q\in\mathbbm{Z}}\mathcal{L}_B^q[\rho]\nonumber\\
\mathcal{L}_B^q[\rho]&=& P_q \big[G_B(\Omega_B + q\nu)\mathcal{D}(A_{\rm B})[\rho] \non\\ 
&+& G_B(-\Omega_B -q\nu)\mathcal{D}(A_{\rm B}^{\dagger})[\rho]\big],
\end{eqnarray}
in terms of the dissipater 
\begin{eqnarray}\label{dissipator}
\mathcal{D}(A_{\alpha})[\rho]=A_{\alpha}\rho A_{\alpha}^\dag -\frac{1}{2}\{\rho,A_{\alpha}^\dag  A_{\alpha}\}.
\end{eqnarray}
 
The operator $A_{\alpha}$ assumes the form $\ket{i}\bra{j}$ ($i \neq j$; $i, j = 1, 2, \hdots, 8$), and causes the twelve transitions described above with positive frequency $\Omega_\alpha=\omega^{\alpha}_{ij}$, between the eigenstates $\ket{i}$ and $\ket{j}$ with frequency difference $\omega_{ij}$, subject to the constraint of four allowed transitions corresponding to the terminal $\alpha$. 
Here $q \in  \mathbbm{Z}$ denotes the different Floquet modes with frequencies $\Omega_B + q\nu$, and amplitudes $P_q$ ($P_q = P_{-q}, \sum_{q \mathbbm{Z}} P_q = 1$). $G_{\alpha}(\omega)=G_0(\omega)(1+\bar{n}_\alpha(\omega))$ is the spectral function of the bosonic bath coupled to the terminal $\alpha$, where $G_0(\omega)$, is the spontaneous-emission rate and $\bar{n}_\alpha(\omega)=\left[\exp{\frac{\hbar\omega}{k_B T_{\alpha}}}-1\right]^{-1}$ is the thermal population of the
bath mode at frequency $\omega$ and temperature $T_{\alpha}$ \cite{alicki2012periodically, kosloff2013quantum, kosloff2014quantum, alicki2014quantum, gelbwaser2015thermodynamics}. 

The Floquet-Lindblad master Eq.~\eqref{master_eq} shows that the TLS base system effectively acts as a multi-level system, with the different energy gaps given by the Floquet modes $q$. The master Eq.~\eqref{master_eq} drives the system toward a Gibbs-like steady state, characterized by the different effective energy gaps, the amplitudes $P_q$, and the spectral functions and temperatures of the baths (see App.~\ref{Appendix-A}).

The steady-state $\rho = \rho_{\rm ss}$, defined by the condition $\dot{\rho}_{ss}=0$, is diagonal in the energy eigenbasis $\ket{j}$ ($j = 1, 2, ... 8$), such that Eq.~\eqref{master_eq} reduces to
\begin{eqnarray}\label{rho-ii-dot}
\dot{\rho}_{11} & = & 0 = \Gamma_{51}^E  + \tilde{\Gamma}_{31}^B + \Gamma_{21}^C, \nonumber  \\
\dot{\rho}_{22} & = & 0 = \Gamma_{62}^E + \tilde{\Gamma}_{42}^B + \Gamma_{12}^C,\nonumber \\
\dot{\rho}_{33} & = & 0 = \Gamma_{73}^E + \tilde{\Gamma}_{13}^B + \Gamma_{43}^C, \nonumber \\
\dot{\rho}_{44} & = & 0 = \Gamma_{84}^E + \tilde{\Gamma}_{24}^B + \Gamma_{34}^C, \nonumber \\
\dot{\rho}_{55} & = & 0 = \Gamma_{15}^E + \tilde{\Gamma}_{75}^B + \Gamma_{65}^C,  \\
\dot{\rho}_{66} & = & 0 = \Gamma_{26}^E + \tilde{\Gamma}_{86}^B + \Gamma_{56}^C, \nonumber \\
\dot{\rho}_{77} & = & 0 = \Gamma_{37}^E + \tilde{\Gamma}_{57}^B + \Gamma_{87}^C, \nonumber \\
\dot{\rho}_{88} & = & 0 = \Gamma_{48}^E + \tilde{\Gamma}_{68}^B + \Gamma_{78}^C. \nonumber
\end{eqnarray}
Here the net transition rates are defined by the following terms:
\begin{eqnarray}\label{gamma_EC}
 \Gamma_{ij}^{\alpha} &&=G_{\alpha} (\omega_{ij}) \rho_{ii} - G_{\alpha} (-\omega_{ij}) \rho_{jj}, \; \alpha \in \{E,C\}; \nonumber \\
\label{gamma_B}
\tilde{\Gamma}_{ij}^{B}  &&=\sum_q \Gamma_{ij,q}^{B}
\end{eqnarray}
where,
\begin{eqnarray}
\Gamma_{ij,q}^{B}&=&P_q[G_{B} (\omega_{ij}+q\nu) \rho_{ii} - G_{B} (-\omega_{ij}-q\nu) \rho_{jj}]\nonumber\\
\Gamma_{ij}^{E(C)}  
 &=& - \Gamma^{E(C)}_{ji}; \quad \tilde{\Gamma}_{ij}^{B}  
 = - \tilde{\Gamma}^{B}_{ji}.
\end{eqnarray}
For simplicity, here, we choose the baths to be Ohmic, so the spectral functions are linear with $G_0(\omega)=\kappa \omega$, where the constant $\kappa$ is the same for all three reservoirs.

Finally, one can use the Spohn inequality~\cite{spohn1978entropy} and the dynamical version of the second law of thermodynamics \cite{kosloff2013quantum} to arrive at the explicit expressions for the steady state heat currents (see App. \ref{Appendix-B}):  
\begin{eqnarray}
J_{E(C)}&=& -\hbar\sum_{\omega_{ij}} \omega_{ij}\Gamma_{ij}^{E(C)} \label{J_EC} \non\\
J_B &=& -\hbar\sum_q \sum_{\omega_{ij}} (\omega_{ij}+q\nu) {\Gamma}_{ij,q}^B .\label{J_B}
\end{eqnarray}

\section{Transistor Characteristics}
\label{Sec.III}

\subsection{General Results}
\label{Sec.IIIA}

The gain of a thermal transistor can be quantified through dynamical amplification parameters:
\begin{equation}\label{gain-beta-bold}
\beta_+ = \frac{\partial J_{C}}{\partial J_B},~~~\beta_- = \frac{\partial J_{E}}{\partial J_B}.
\end{equation} 
Large values of $\beta_{\pm}$ imply high amplifications, which may even diverge for $J_B$ passing through a minimum. 
We note that conservation of energy demands $J_B+J_E+J_C=0$, which in turn results in the relation $\beta_+ + \beta_- = -1$; here $J_E>0$ and $J_B, J_C <0$ in resemblance with the common base electronic transistor~\cite{millman2009millman}.

\begin{figure}[!]
\centering	\includegraphics[width=0.8\columnwidth]{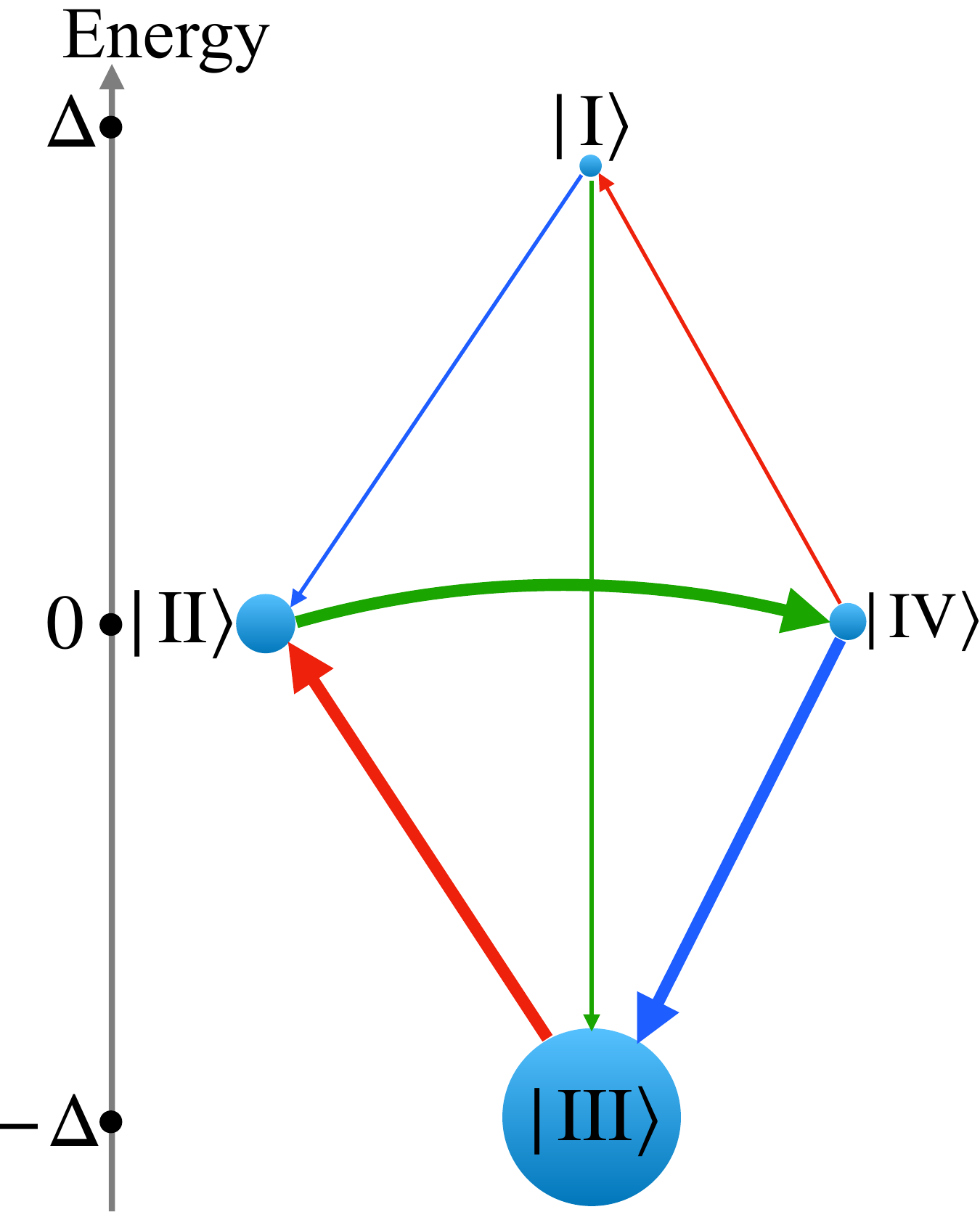}
\caption{Energy levels for $\omega_E=\omega_0=\omega_C=0$, $\omega_{EC}=0$, and $\omega_{EB}=\omega_{BC}=\Delta$. The arrows indicate the net decaying rats $\Gamma^{\alpha}_{i-j}$ between the states $\left|I\right>$, $\left|II\right>$, $\left|III\right>$  and $\left|IV\right>$, due to bath $E$ (red), bath $B$ (green), and bath $C$ (blue).}
\label{FQT_fig2}
\end{figure}

Let us now look into the details of the dynamics that allow us to operate the FQT with high $\beta_{\pm}$. In order to devise a common-base FQT, we set the emitter-collector coupling $\omega_{EC}$ to be zero, while the emitter-base and base-collector couplings are taken to be non-zero and equal, i.e., $\omega_{EB} = \omega_{BC} = \Delta$ and (see Eq. \eqref{model-Hamiltonian} and App.~\ref{Appendix-C}). Note that if coupling between all three qubits are equal (symmetric) then the transistor effect disappears. Although one may achieve thermal transistor effect with asymmetric couplings and non-zero qubit frequencies as well (for details see Ref.~\cite{joulain2016quantum}). However, in order to reduce the number of the states and avoid numerical complexity, we set bare frequencies of all the TLSs as zero, i.e., $\omega_{C} = \omega_{E} = \omega_0 = 0$. Under this choice of parameters, some of the eigenstates become degenerate, so that we are finally left with only 3 distinct energy levels. We rename the states $|1\rangle$ and $|8\rangle$ as $|I\rangle$, $|2\rangle$ and $|7\rangle$ as $|II\rangle$, $|3\rangle$ and $|6\rangle$ as $|III\rangle$ and $|4\rangle$ and $|5\rangle$ as $|IV\rangle$ [Fig.~\ref{FQT_fig2}]. Similarly, we introduce new density matrix elements as $\rho_{I}=\rho_{11}+\rho_{88}$, $\rho_{II}=\rho_{22}+\rho_{77}$, $\rho_{III}=\rho_{33}+\rho_{66}$, and $\rho_{IV}=\rho_{44}+\rho_{55}$. Introducing the net decaying rates $\Gamma_{i - j}^{\alpha}$ ($\alpha = {E, B, C}$; $i, j = I, II, III, IV$) between these states (see App. \ref{Appendix-C}) following Eq.~\eqref{J_B}, the three currents  can be written as
\begin{eqnarray}\label{hc-expression}
&& J_E= -\hbar\Delta \left[\Gamma_{I-IV}^E + \Gamma_{II-III}^E\right], \nonumber \\
&& J_B = -\hbar \sum_{q=0,\pm 1} (2 \Delta + q \nu) \Gamma_{I-III,q}^B, \nonumber \\
&& J_C = -\hbar\Delta \left[ \Gamma_{I-II}^C  + \Gamma_{IV-III}^C \right].
\end{eqnarray}
where
\begin{eqnarray}\label{decay_rates-EC}
&& \Gamma_{I-IV}^E =    \kappa \Delta \{ \rho_{I}-e^{-\hbar\Delta/k_BT_{ E}} \rho_{IV} \} \nonumber\\
&& \Gamma_{II-III}^E =    \kappa \Delta \{ \rho_{II}-e^{-\hbar\Delta/k_BT_{ E}} \rho_{III} \}\nonumber \\
&&  \Gamma_{IV-III}^C =    \kappa \Delta \{ \rho_{IV}-e^{-\hbar\Delta/k_BT_{ C}} \rho_{III} \}\nonumber \\
 &&  \Gamma_{I-II}^C  =    \kappa \Delta \{ \rho_{I}-e^{-\hbar\Delta/k_BT_{ C}} \rho_{II} \}\nonumber\\
\Gamma_{I-III,q}^B &&= P_q (2\Delta +q\nu) \{\rho_{I}-e^{-\hbar(2\Delta+q \nu)/k_B T_{ B}}\rho_{III}\}\nonumber\\
\end{eqnarray}

To ensure that the system dynamics fulfils Born-Markov approximations, the system relaxation time $\Gamma^{-1}$ must be larger than the characteristic time scale associated with the system frequency, i.e., $\Delta^{-1}$. As we show in App.~\ref{Appendix-C}, level decaying rates exhibit significant enhancement with the increase of base temperature, which indicates that the Born-Markov approximation may no longer be a good approximation at high temperature limits. As a result, we choose the highest temperature $T_{ E}$ such that $e^{-\hbar\Delta/k_BT_{ E}} \ll 1$.   

Transitions between the different states of FQT are illustrated in Fig.~\ref{FQT_fig2}. The arrows denote the transition directions whereas larger widths indicate higher decay rates $\Gamma^{\alpha}_{i-j}$.  As shown in Fig.~\ref{FQT_fig2}, the energy exchanges are primarily dominated by the $III-II$ and $IV-III$ transitions. Consequently, one expects $J_E, J_C \gg J_B$ (see Eq. \eqref{hc-expression}). These observations can be explained by examining carefully the following populations and heat current expressions. If we limit our calculation in the leading orders of $e^{-\hbar\Delta/k_B T_{ E}}$ and the first two harmonics of the modulation (i.e., $|q| = 0, 1$),  one can evaluate the approximate form of populations (see App. \ref{Appendix-C} for detailed derivation) as 
\begin{eqnarray}
\rho_I &&\simeq \frac{1}{2(1+P_0+2P_1)} \nonumber\\
&&\times \frac{P_0 k_B T_{ B} + P_1 \hbar\nu \left(\frac{e^{\hbar\nu/{k_B T_{ B}}}+1}{e^{\hbar\nu/{k_B T_{ B}}}-1}\right)}{2P_0 k_B T_{B} + 2P_1 \hbar\nu \left(\frac{e^{\hbar\nu/{k_B T_{ B}}}+1}{e^{\hbar\nu/{k_B T_{ B}}}-1}\right)+\hbar\Delta}e^{-2\hbar\Delta/{k_B T_{ E}}},\nonumber\\
\rho_{II} &&\simeq  \frac{P_0 k_B T_{B} + P_1\hbar \nu \left(\frac{e^{\hbar\nu/{k_B T_{ B}}}+1}{e^{\hbar\nu/{k_B T_{ B}}}-1}\right)+\hbar\Delta}{2P_0 k_B T_{ B} + 2P_1 \hbar\nu \left(\frac{e^{\hbar\nu/{k_B T_{B}}}+1}{e^{\hbar\nu/{k_B T_{ B}}}-1}\right)+\hbar\Delta}e^{-\hbar\Delta/{k_B T_{ E}}},\nonumber\\
\rho_{III} &&\simeq 1- e^{-\hbar\Delta/{k_B T_{ E}}},\nonumber\\
\rho_{IV} && \simeq  \frac{P_0 k_B T_{B} + P_1 \hbar\nu \left(\frac{e^{\hbar\nu/{k_B T_{B}}}+1}{e^{\hbar\nu/{k_B T_{ B}}}-1}\right)}{2P_0 k_B T_{ B} + 2P_1 \hbar\nu \left(\frac{e^{\hbar\nu/{k_B T_{B}}}+1}{e^{\hbar\nu/{k_B T_{ B}}}-1}\right)+\hbar\Delta}e^{-\hbar\Delta/{k_B T_{ E}}}.\nonumber\\\label{population-gen}
\end{eqnarray}
To obtain the above set of closed form analytical expressions, we take for simplicity $T_{C}, T_{ B} \ll T_{E}$ so that one can neglect the contribution of $e^{-\hbar\Delta/k_BT_{ C}}$ and $e^{-\hbar\Delta/k_BT_{B}}$ w.r.t $e^{-\hbar\Delta/k_BT_{ E}}$ in the populations expressions. Using Eqs.~\eqref{hc-expression}-\eqref{population-gen}, we now explicitly evaluate the three thermal currents as 
\begin{eqnarray}
J_E &\simeq& -J_C \simeq \kappa\hbar \Delta^2 \rho_{IV}   \non\\
J_B &\simeq& \kappa\hbar \sum_{q=0,\pm 1} P_q(2 \Delta + q \nu)^2 \Big[e^{-\hbar(2\Delta+q \nu)/{k_B T_{ B}}} - \rho_I\Big].\nonumber\\
\label{heat-currents-gen}
\end{eqnarray}

One notes that $J_E$ and $J_C$ are driven by $\rho_{IV}$, i.e., the state populations at the intermediate energy levels. On the other hand, $J_B$ is determined by the population $\rho_I$ of the most energetic states, with all spins aligned parallel.  As one can see from Eq. \eqref{population-gen}, we have $\rho_I \ll \rho_{IV}$ in the limit of $k_{B} T_{ E} \ll \hbar \Delta$  (see App. \ref{Appendix-C}). This in turn leads to  $J_B \ll J_{E}, J_{C}$. Furthermore, as seen from Eqs. \eqref{population-gen} and \eqref{heat-currents-gen}, one can control the populations and consequently the heat currents through control of the modulation frequency $\nu$. In what follows, we consider different choices of the parameters that allow us to operate the FQT in the following three regimes discussed below.

\subsection{Unmodulated quantum thermal transistor}\label{Sec.IIIB}

In absence of modulation, i.e., upon setting the limit $\nu \rightarrow 0$ and taking $P_q = \delta_{q,0}$ in Eqs.~\eqref{population-gen}-\eqref{heat-currents-gen}, we qualitatively reproduce the results reported in Ref. \cite{joulain2016quantum} for the populations and heat currents. Throughout our calculation we have chosen $\kappa=1$.  In this limit we get 
\begin{eqnarray}
\rho_I &&\simeq  \frac{k_B T_{ B}}{8k_B T_{B} + 4\hbar\Delta}e^{-2\hbar\Delta/{k_B T_{ E}}},\nonumber\\
\rho_{II} &&\simeq  \frac{k_B T_{ B} +\hbar\Delta}{2k_B T_{ B} + \hbar\Delta}e^{-\hbar\Delta/{k_B T_{ E}}},\nonumber\\
\rho_{III} &&\simeq 1- e^{-\hbar\Delta/{k_B T_{ E}}},\nonumber\\
\rho_{IV} && \simeq \frac{k_B T_{B}}{2k_B T_{ B} + \hbar\Delta}e^{-\hbar\Delta/{k_B T_{E}}},
\label{population-I}
\end{eqnarray}
and 
\begin{eqnarray}
&& J_E \simeq -J_C \simeq \kappa\hbar
\Delta^2 \rho_{IV} \nonumber \\
&& J_B \simeq 4\kappa\hbar\Delta^2 \left[e^{-2\hbar\Delta/{k_B T_{ B}}}-\rho_I\right]. 
\label{hc-QTT}
\label{heat-currents-I}
\end{eqnarray}
\begin{figure}
\centering
\includegraphics[width=\columnwidth]{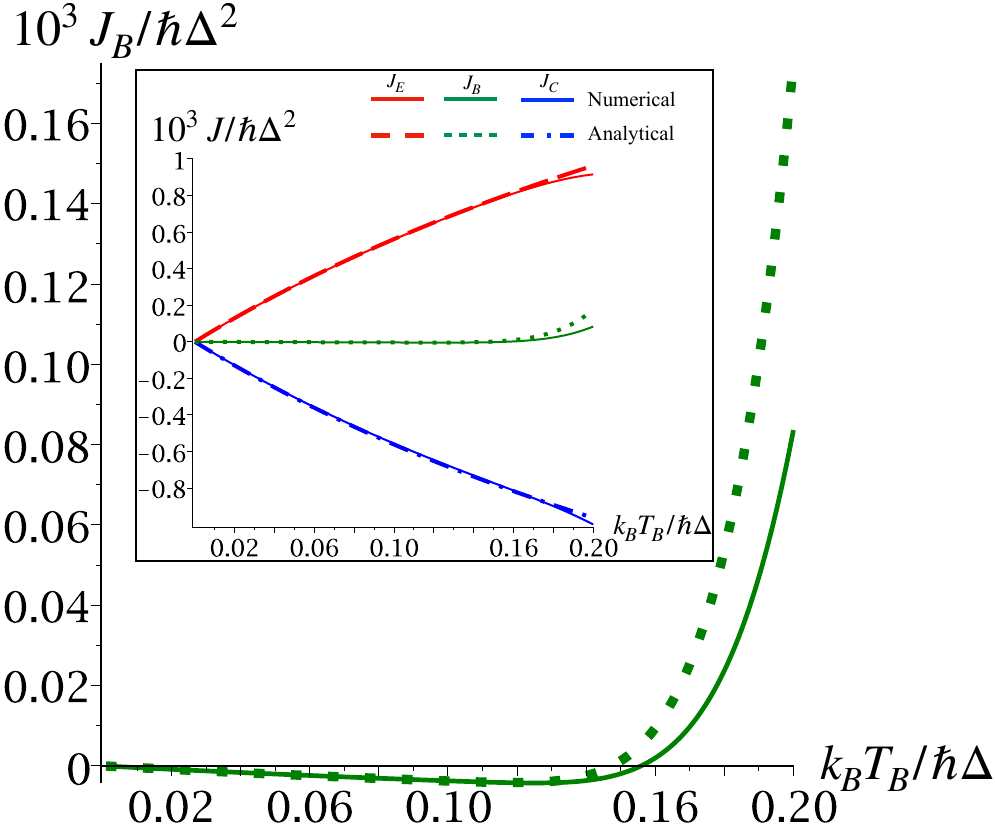}
\caption{Thermal current $J_B$ versus $k_BT_{ B}/\hbar\Delta$ for the unmodulated case. Solid lines indicate numerical results, while dotted lines are approximate analytical results [Cf. Eq.~\eqref{hc-QTT}]. Inset: thermal currents $J_E$, $J_B$, and $J_C$  versus $k_BT_{ B}/\hbar\Delta$ for unmodulated case with parameters, $\omega_E=\omega_0=\omega_C=0$, $\omega_{EC}=0$, $\omega_{EB}=\omega_{BC}=\Delta$, $k_B T_{ E}=0.2\hbar \Delta$, and $k_B T_{ C}=0.02\hbar \Delta$. }
\label{FQT_fig3}
\end{figure}
\begin{figure}
\centering
\includegraphics[width=\columnwidth]{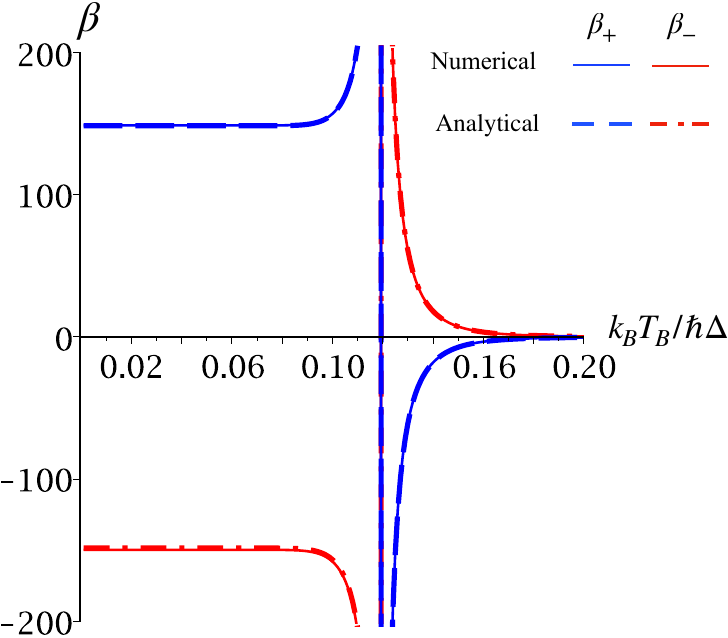}
\caption{Dynamical amplification factors [Cf.~\eqref{gain-beta-bold}] $\beta_+$ and $\beta_-$ versus $k_B T_{ B}/\hbar\Delta$ for the unmodulated case with parameters, $\omega_E=\omega_0=\omega_C=0$, $\omega_{CE}=0$, $\omega_{EB}=\omega_{BC}=\Delta$, $k
_BT_{ E}=0.2\hbar\Delta$, and $k
_BT_{ C}=0.02\hbar\Delta$. }
\label{FQT_fig4}
\end{figure}
 Apart from a negligible contribution of the order of $e^{-2\hbar\Delta/{k_B T_{B}}}$, the populations and heat currents in Eqs. \eqref{population-I} and \eqref{hc-QTT} are identical to those reported in Ref.~\cite{joulain2016quantum}, and they match the numerical results exactly, except for $T_{B} \to T_{E}$ (see Fig.~\ref{FQT_fig3} and App. \ref{Appendix-C}). As expected, $|J_B| \ll |J_E|,~|J_C|$; a small change of $J_B$ can  significantly alter the values of $J_E$ and $J_C$, thus resulting in large values of the  dynamical amplification factors $\beta_+$ and $\beta_-$, as  shown in Fig. \ref{FQT_fig4}. From Eq.~\eqref{hc-QTT}, one can estimate that as long as $T_{ B}$ is not comparable to $T_{E}$, the absolute value of the dynamical amplification factors are approximately $|\beta_{\pm}| \approx e^{\hbar \Delta/k_{B} T_{ E}}$. For the present choice of system parameters $\hbar\Delta/k_B T_{ E}=5$, the setup yields a transistor gain $\beta_{\pm} \sim e^{5} \approx 150$ [Fig.~\ref{FQT_fig4}], which corroborates with our numerical findings. Finally, we note that $J_{C}$, $J_{E}$ and $J_{B}$ all tend to vanish in the limit $T_{B} \rightarrow 0$, so that the transistor remains in the so called {\it cut-off} regime, and the device stops operating as a viable heat current modulator.

\subsection{Sinusoidal and $\pi$-flip modulations}\label{Sec.IIIC}
\begin{figure}[!]
\centering
\includegraphics[width=0.9\columnwidth]{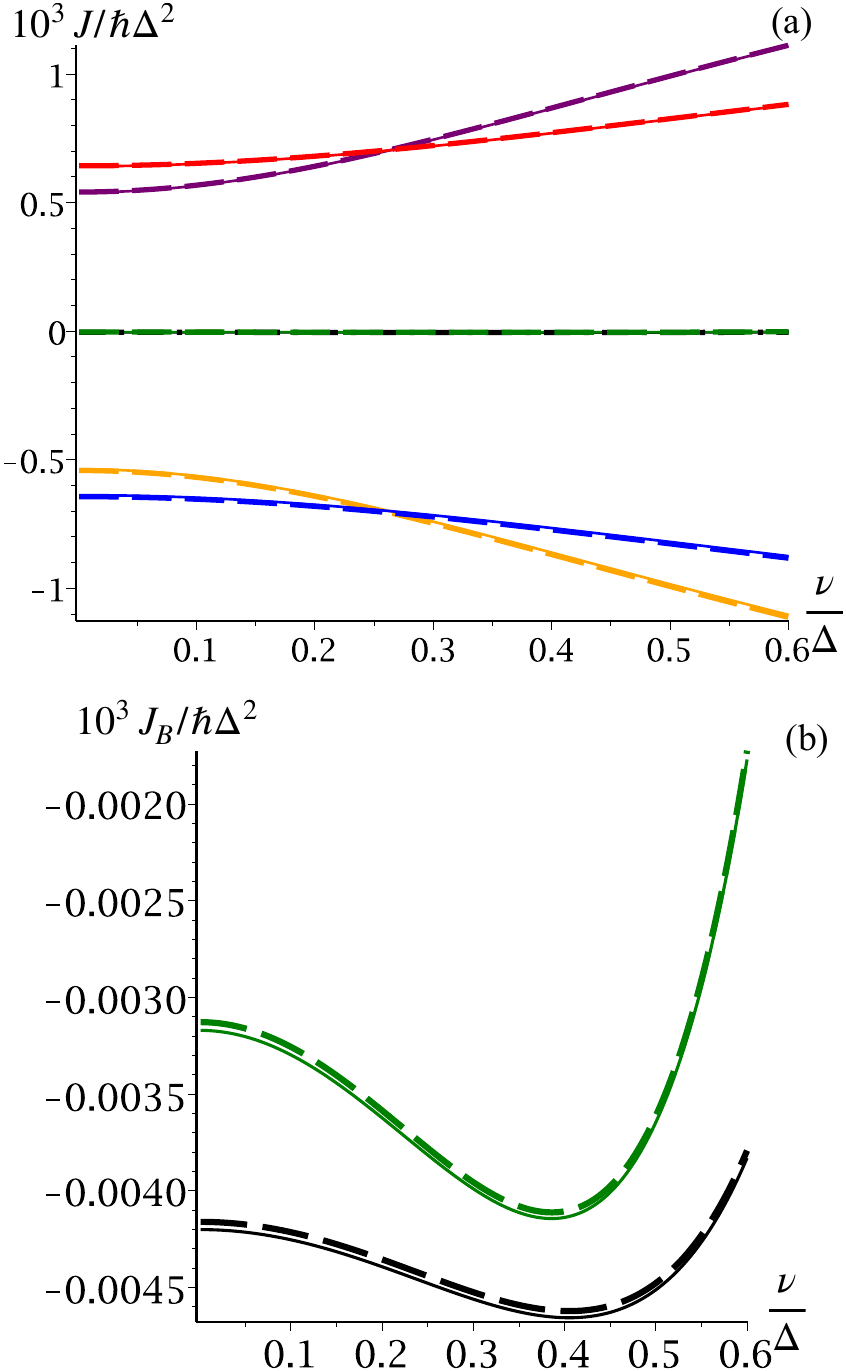}
\caption{(a) Thermal currents $J_E$, $J_B$, and $J_C$ versus $\nu/\Delta$ under sinusoidal modulation [$J_E$ (red), $J_B$ (black) and $J_C$ (blue)] and $\pi$-flip modulation [$J_E$ (purple), $J_B$ (green) and $J_C$ (orange)].(b) Thermal current $J_B$ versus $\nu/\Delta$ for sinusoidal (black) and $\pi$-flip modulation (green). Solid lines indicate numerical values and dashed lines are approximate analytical values of heat currents for the parameter set: $\omega_E=\omega_0=\omega_C=0$, $\omega_{EC}=0$, $\omega_{EB}=\omega_{BC}=\Delta$, $k_BT_{ E}=0.2\hbar\Delta$, $k_BT_{ B}=0.118\hbar\Delta$, $k_BT_{ C}=0.02\hbar\Delta$ and $\lambda=0.8$. }
\label{FQT_fig5}
\end{figure}
We now study the operation of the FQT in presence of periodic modulations of $\omega_{ B}(t)$. In particular, we consider sinusoidal and  $\pi$-flip modulations, as discussed below.

{\bf Sinusoidal modulation:} Here we consider the modulation form
 \begin{equation}
 \omega_{B} (t) = \omega_0 + \lambda \nu \sin (\nu t ).
 \end{equation}
The condition $0 \le \lambda \le 1$ allows us to limit our analysis to only the first two harmonics $q=0, \pm 1$ in Eqs. \eqref{population-gen} and \eqref{heat-currents-gen} \cite{gelbwaser2013minimal}:
\begin{equation}
    P_0 = 1-\frac{{\lambda}^2}{2}, \quad P_{\pm}1 = \frac{{\lambda}^2}{4}.
\end{equation}
\begin{figure}[!]
\centering
\includegraphics[width=\columnwidth]{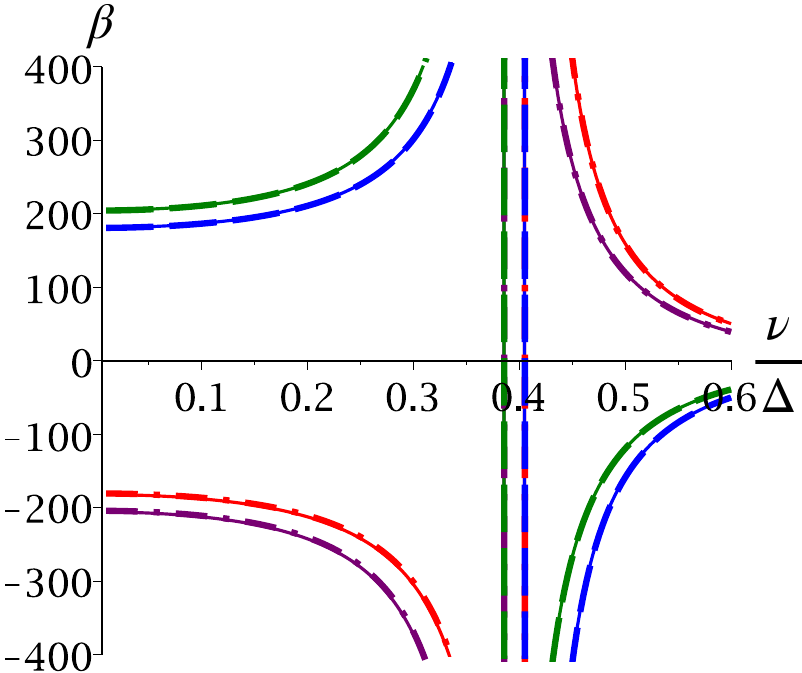}
\caption{Dynamical amplification factors [Cf.~\eqref{gain-beta-bold}] $\beta_+$ and $\beta_-$ versus $\nu/\Delta$ for sinusoidal [$\beta_+$ (blue) and $\beta_-$ (red)] and $\pi$-flip modulation [$\beta_+$ (green) and $\beta_-$ (purple)]. Solid lines show numerical results, while dashed-dot and dashed lines correspond to approximate analytical value of $\beta_{\pm}$ for the parameter set $\omega_E=\omega_0=\omega_C=0$, $\omega_{CE}=0$, $\omega_{EB}=\omega_{BC}=\Delta$, $k_BT_{ E}=0.2\hbar\Delta$, $k_BT_{ C}=0.02\hbar\Delta$, $k_BT_{ B}=0.118\hbar\Delta$ and $\lambda=0.8$. }
\label{FQT_fig6}
\end{figure}

{\bf $\pi-$flip modulation:} Periodic $\pi$- phase shifts modulation in the form of two alternating pulses per period~\cite{alicki2012periodically}
\begin{eqnarray}
\omega(t)=\omega_0+\pi\sum_{n \in \mathbbm{Z}} \delta(t-(n+1/4))-\delta(t-(n+3/4)),\nonumber\\
\end{eqnarray}
gives rise to only two leading harmonics $q = \pm 1$, with the corresponding amplitudes of the Floquet modes given by \cite{gelbwaser2013minimal};
 \begin{equation}
     P_{\pm 1} \approx (2/\pi)^2.
     \label{Ppi}
 \end{equation}
In this case $P_0 =0$. The results for sinusoidal and $\pi$-flip modulations are shown in Figs. \ref{FQT_fig5}. As one can infer from Eqs. \eqref{population-gen} and \eqref{heat-currents-gen}, in comparison to sinusoidal modulation, the vanishing $P_0$ for pi-pulse  results in larger values of $\beta_{\pm}$ due to higher slopes in the heat currents (Figs.~\ref{FQT_fig5} and \ref{FQT_fig6}).   

\subsection{Generic modulation with $T_{\rm B} \to 0$}
\label{Sec.IIID}

\begin{figure}[!]
\centering
\includegraphics[width=\columnwidth]{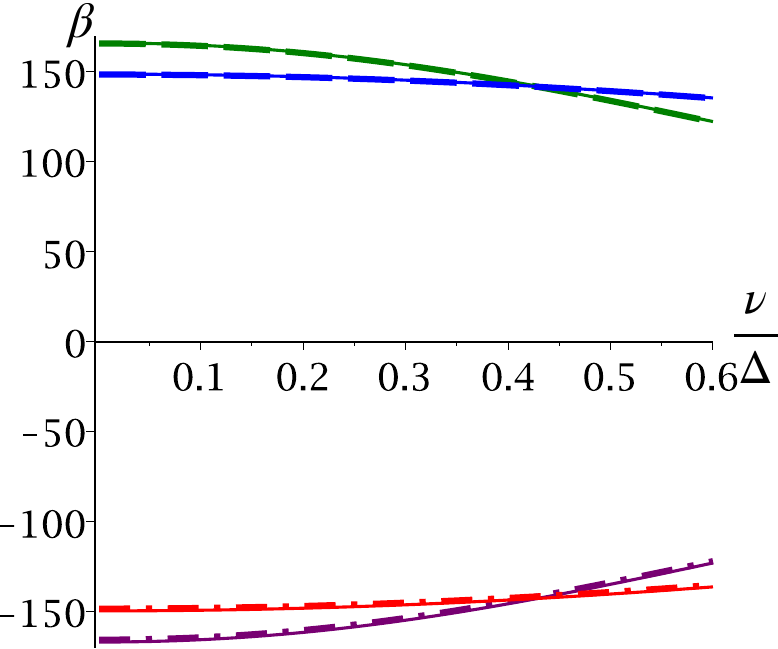}
\caption{Dynamical amplification factors [Cf.~\eqref{gain-beta-bold}] $\beta_+$ and $\beta_-$ versus $\nu/\Delta$ for sinusoidal [$\beta_+$ (blue) and $\beta_-$ (red)] and $\pi$-flip modulations [$\beta_+$ (green) and $\beta_-$ (purple), for $T_{ B} \to 0$. Solid lines represent numerical values, dashed-dot and dashed lines correspond to approximate analytical values of $\beta_{\pm}$ for the parameter set $\omega_E=\omega_0=\omega_C=0$, $\omega_{CE}=0$, $\omega_{EB}=\omega_{BC}=\Delta$, $k_BT_{ E}=0.2\hbar\Delta$, $k_BT_{C}=0.02\hbar\Delta$ and $\lambda=0.8$. }
\label{FQT_fig8}
\end{figure}

We now emphasize the crucial advantage offered by the FQT model proposed here. As we show below, in contrast to the previously studied models of quantum thermal transistors \cite{zhang2018coulomb,
guo2019multifunctional,
mandarino2021thermal,
joulain2016quantum}, FQT can operate as a thermal current amplifying device in the otherwise cut-off regime of the base temperature $T_{ B}$ approaching zero. In this limit the populations and heat currents of the FQT reduce to (see Eqs.~\eqref{population-gen}-\eqref{heat-currents-gen})
\begin{figure}[!]
\centering
\includegraphics[width=0.95\columnwidth, height=13.8 cm]{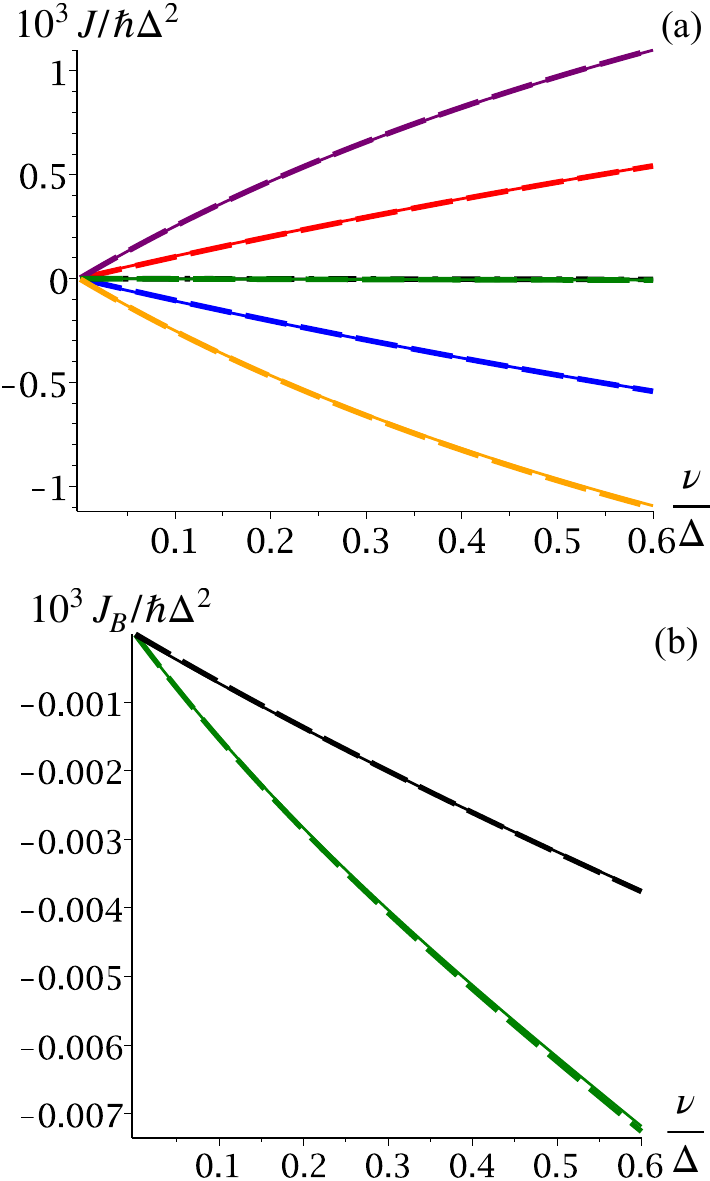}
\caption{(a) Thermal currents $J_E$, $J_B$, and $J_C$ versus $\nu/\Delta$ under sinusoidal modulation [$J_E$ (red), $J_B$ (black) and $J_C$ (blue)] and $\pi$-flip modulation [$J_E$ (purple), $J_B$ (green) and $J_C$ (orange)],(b)Thermal current $J_B$ versus $\nu/\Delta$ for sinusoidal (black) and $\pi$-flip modulations (green), for $T_{ B} \to 0$. Solid lines represent numerical values and dashed lines stand for approximate analytical values of heat currents for the parameter set $\omega_E=\omega_0=\omega_C=0$, $\omega_{EC}=0$, $\omega_{EB}=\omega_{BC}=\Delta$, $k_BT_{ E}=0.2\hbar\Delta$, $k_BT_{ C}=0.02\hbar\Delta$ and $\lambda=0.8$. }
\label{FQT_fig7}
\end{figure}
\begin{eqnarray}
\rho_I &&\simeq \frac{1}{2(1+P_0+2P_1)} \frac{ P_1 \nu }{2P_1 \nu +\Delta}e^{-2\hbar\Delta/{k_B T_{ E}}}, \non\\
\rho_{II} &&\simeq  \frac{ P_1 \nu + \Delta }{2P_1 \nu + \Delta}e^{-\hbar\Delta/{k_B T_{ E}}},\non\\
\rho_{III} &&\simeq 1- e^{-\hbar\Delta/{k_B T_{ E}}},\non\\
\rho_{IV} && \simeq  \frac{ P_1 \nu }{2P_1 \nu + \Delta}e^{-\hbar\Delta/{k_B T_{ E}}}
\end{eqnarray}
and 
\begin{eqnarray}
J_E &\simeq& -J_C \simeq \kappa\hbar \Delta^2 \rho_{IV} \non\\
J_B &\simeq& -\kappa\hbar \rho_{I}\sum_{q=0,\pm 1}  P_q(2 \Delta + q \nu)^2.
\label{JCJB_TB0}
\end{eqnarray}
Clearly the non zero Floquet amplitude $P_{1}$, which is a direct consequence of the modulation, results in non-zero $\rho_I$, and therefore allows one to operate the FQT with non-zero heat currents even  for $T_{\rm B} \to 0$.
As before, we again focus on sinusoidal and $\pi$-flip modulations to exemplify the advantage offered by the FQT.
We plot the heat currents for sinusoidal and $\pi$-flip modulations in Fig.~\ref{FQT_fig7}; As expected from Eq. \eqref{JCJB_TB0}, sinusoidal and $\pi$-flip modulations result in large amplification factors even in this regime [Fig. \ref{FQT_fig8}]. 

\section{Conclusion}
\label{secIV}

We have proposed operating quantum thermal transistors  in presence of periodic modulations, and termed the setup as Floquet Quantum Thermal Transistors. In previously studied quantum thermal transistors,  transistor effect was realized through changes in temperature $T_{ B}$ of the bath coupled to the base \cite{joulain2016quantum, ben2014near, li2006negative, guo2018quantum, zhang2018coulomb, ghosh2021quantum, mandarino2021thermal, naseem2020minimal}. This can be expected to be an energy-intensive process; a change $\delta T_{ B}$ in temperature would require a heat energy $\delta Q_{ B} = C_{B} \delta T_{ B}$, which can be large for a large heat capacity $C_{B}$ of the bath. On the other hand, 
the periodic modulation scheme implemented here allows one to modulate the heat currents also in presence of fixed bath temperatures. We note that the cost of periodic modulation is an interesting question as well, which we plan to address in future works. We have exemplified our generic theory using sinusoidal and $\pi$-flip modulations.   The crucial advantage of the FQT is its ability to operate as a heat modulation device with high amplification factors even in the traditional cut-off regime of $T_{B} \to 0$. This advantage stems from non-zero floquet amplitudes $P_1$, which vanishes in absence of modulation. Although we use Ohmic bath spectral density throughout our calculation, one may expect transistor effect yet in presence of non-Ohmic bath spectrum, (such as Lorentzian) as long as basic energy level diagram (Fig.~\ref{FQT_fig2}) characterized by net decaying rates fulfills $J_B<<J_E,J_C$. However a complete analysis in presence of a generic non-Ohmic bath is non-trivial and beyond the scope of present work.

Experimental realization of controlled heat currents can be envisaged in a variety of physical platforms: (i) quantum thermal transistor design has been predicted and realized in on-chip superconducting qubits \cite{karimi2016otto}, (ii) quantum heat engine has been realized with the spin-1/2 of ${}^{13}\text{C}$ nucleus \cite{batalhao2014experimental}, (iii) micro- and nano-electromechanical Systems (MEMS/NEMS) have been used to implement nonequilibrium reservoir engineering~\cite{klaers2017squeezed} etc. Most of these  quantum technology devices require sohphisticated lithographic techniques (few tens of nanometers) and low temperature measurements. In our approach, experimental feasibility relies on the periodic modulation of the base bath (ultra-thin film) in conventional electronic bipolar junction transistor (BJT) architecture. Recent analytical results on far-field radiative heat transport concerns thermal transistor (BJT) set-up where 1 ${\mu}${\rm m}-thick block of vanadium dioxide (${\rm VO_2}$), as base, has been excited with a laser with a modulation frequency of 0.5 Hz \cite{ordonez2016dynamical, ordonez2019radiative}. Interestingly, thermal rectification has been achieved due to the insulator-to-metal transition of ${\rm VO_2}$ at around 340 K, which can thus set the temperature scale of a typical base bath. As reported in Ref~\cite{ordonez2016dynamical}, with practical design, values for amplification factor close to 10 can be achieved choosing the optimal frequency of modulation ($\sim$ 1 Hz). With these paramerters used in the experimental setup~\cite{ordonez2016dynamical}, the modulation scheme presented here may lead to experimentally observable heat fluxes with higher amplitudes.

\section*{ACKNOWLEDGEMENTS}

N.G. and S.B. thank, respectively, CSIR (SRF) and DST INSPIRE for financial support through fellowships. A.G. is thankful for Initiation Grant, IITK (Grant No. IITK/CHM/2018513) and SRG (Grant No. SRG/2019/00028), SERB, India for the partial financial support.
B.D. is grateful to IACS for the fellowship. S.D. acknowledges the financial support from DST-SERB grant nos. ECR/2017/002037. S.D. also acknowledges support from the Technical Research Centre (TRC), IACS, Kolkata.
V.M. acknowledges support from Science and Engineering Research Board (SERB) through MATRICS (Project No.
MTR/2021/000055) and Seed Grant from IISER Berhampur.

\onecolumngrid 

\appendix

\section{Derivation of the Floquet-Master Equation}\label{Appendix-A}
We start with interaction Hamiltonian 
\ba
&&H_I=H_{IE}+H_{IB}+H_{IC}=\sigma_{x}^{E}\otimes \mathcal{R}_{E} + \sigma_{x}^{B}\otimes \mathcal{R}_{B} + \sigma_{x}^{C}\otimes \mathcal{R}_{C},\label{HI-Hamiltonain}
\ea
and in terms of the time ordered unitary operator
\ba
U(t,0) = \mathcal{T}\exp\left(-\frac{i}{\hbar}\int^{t}_{0} H_S(t^{\prime}) dt^{\prime} \right),
\label{U(t,0)}
\ea
we obtain interaction picture von-Neumann equation for the total density matrix $\rho_T(t)$ of the combined system as
\ba
&&\frac{d}{dt}\rho_T(t) = -\frac{i}{\hbar} \left[H_{I}(t),\rho_T(t)\right].
\label{rhoT}
\ea
Integrating the above equation and taking a trace over the bath degrees of freedom we obtain
\ba
&&\frac{d}{dt}\rho_s(t) = -\frac{1}{\hbar^2}\int^{t}_0 ds {\rm Tr}_{E,B,C}\Big[H_{I}(t),\left[H_{I}(s),\rho_T(s)\right]\Big],
\ea
where we use ${\rm Tr}_{E,B,C}\{\rho_T(t)\}=\rho_s(t)$ and assume
${\rm Tr}_{E,B,C}\left[H_I(t),\rho_T(0)\right]=0$. Here ${\rm Tr}_{E,B,C}$ refers to the trace over each bath degrees of freedom. Under Born-Markov approximation, the reduced dynamics of the system in the weak system-bath coupling limit can be written as~\cite{carmichael1999book,breuer2002book,wijesekara2020optically,wijesekara2021darlington,mondal2020exploring} 
\ba
&&\frac{d}{dt}\rho_s(t) = -\frac{1}{\hbar^2}\int^{\infty}_0 ds {\rm Tr}_{E,B,C}\Big[H_{I}(t),\left[H_{I}(t-s),\rho_s(t)\otimes\rho_E\otimes\rho_B\otimes\rho_C\right]\Big],
\label{Born-Morkov-rho_s}
\ea
where we substitute
$\rho_T(t)=\rho_s(t)\otimes\rho_E\otimes\rho_B\otimes\rho_C$. Here the interaction picture Hamiltonian [Cf. Eq.~\eqref{HI-Hamiltonain}]is given by
\ba
H_I(t)&=& \sum_{\alpha \in {\{E,B,C\}}}H_{I\alpha}(t);\nonumber\\
H_{I\alpha}(t)&=& \sigma_{x}^{\alpha}(t)\otimes \mathcal{R}_\alpha(t)=\Big(\sigma_{+}^{\alpha}(t)+\sigma_{-}^{\alpha}(t)\Big)\otimes\sum_{k}\Big(g_k b^{\alpha}_k(t)+g^{*}_k b^{\alpha \dagger}_k(t)\Big)
\label{int-H}
\ea
where $\mathcal{R}_{\alpha}(t)=\sum_{k}(g_k b^{\alpha}_k(t)+g^{*}_k b^{\alpha \dagger}_k(t))$, satisfying 
$\langle{\mathcal{R}_{\alpha}(t)}\rangle= {\rm Tr}_{\alpha}\{\mathcal{R}_{\alpha}(t)\rho_\alpha\}=0;\; \alpha = E,B,C.$  This implies~\cite{wijesekara2020optically,breuer2002book,mondal2020exploring}:
\ba
{\rm Tr}_{E,B,C}\Biggl\{\Big[H_{I\alpha}(t),\left[H_{I\beta}(t-s),\rho_s(t)\otimes\rho_E\otimes\rho_B\otimes\rho_C\right]\Big]\Biggr\}=0; \quad \alpha\ne\beta, \quad \alpha,\beta=E,B,C.
\ea
As a result Eq.~\eqref{Born-Morkov-rho_s} simplifies to
\ba
&&\frac{d}{dt}\rho_s(t) = -\frac{1}{\hbar^2} \sum_{\alpha \in {\{B,C,E\}}}\Biggl\{\int^{\infty}_0 ds {\rm Tr}_{E,B,C}\Big[H_{I\alpha}(t),\left[H_{I\alpha}(t-s),\rho_s(t)\otimes\rho_{E}\otimes\rho_{B}\otimes\rho_{C}\right]\Big]\Biggr\},\label{reduced-master-eqn}
\ea
Now deriving Eq.~\eqref{master_eq} from the above equation is straightforward following the standard procedure for a single thermal reservoir~\cite{breuer2002book},  separately for each $\mathcal{R}_{\alpha}$, while calculation of system operator $\sigma^{\alpha}_x(t)$ requires some elucidations.

For left (emitter) and right (collector) TLS system $\sigma_{\pm}(t)$ can be calculated in a straight forward way:
\ba
\sigma^{\alpha}_{\pm}(t)=U^\dagger(t,0)\sigma^{\alpha}_{\pm}U(t,0)=\exp\left[\frac{i}{\hbar}\int_0^t H_S(s)ds\right]\sigma_{\pm}^{\alpha} \exp\left[-\frac{i}{\hbar}\int_0^t H_S(s)ds\right]; \quad \alpha \in \{E,C\},\label{sigma-ec-t}
\ea
where,
\begin{equation}
     H_S(t)=\frac{\hbar\omega_E}{2}\sigma_z^E+\frac{\hbar\omega_B(t)}{2}\sigma_z^B+\frac{\hbar\omega_C}{2}\sigma_z^C +\frac{\hbar\omega_{EB}}{2}\sigma_z^E\sigma_z^{B} 
+\frac{\hbar\omega_{BC}}{2}\sigma_z^B\sigma_z^{C}+\frac{\hbar\omega_{CE}}{2}\sigma_z^C\sigma_z^{E}.\label{sys-H-s-t}
\end{equation}    
Among the six terms in the each exponent, only three terms will survive for each case; the time-dependent part of $H_S$ also cancels from both side of the exponent. Therefore, Eq.~\eqref{sigma-ec-t} reduces to 
\ba
\sigma^{\alpha}_{\pm}(t)=U^\dagger(t,0)\sigma^{\alpha}_{\pm}U(t,0)=\sum_{\{\Omega_\alpha\}}e^{\pm i\Omega_\alpha t}\sigma^{\alpha}_{\pm}; \quad \alpha \in \{E,C\}\label{sigma_ec_t}
\ea
But for middle (base) TLS, calculation of $\sigma^B_{\pm}(t)$ is nontrivial because of the presence of time dependent factor $\exp(i\omega_B(t) t)$ in $U(t,0)$. However our FQT system Hamiltonian is periodic in $\tau$, $H_S(t + \tau) = H_S(t)$, so we use the Floquet theorem~\cite{floquet1883equations,
mondal2020exploring}to decompose the time evolution operator as $U(t,0) = P(t)e^{R t}$, where $P(t)$ is $\tau$-periodic and $R$ is a constant operator. Because of the periodicity of $P(t)$, it follows from $U(0,0)=\mathbbm{1}$ that $P(0)=\mathbbm{1}$ and hence $U(\tau,0)=P(\tau)e^{R \tau}\equiv P(0)e^{R \tau}=e^{R \tau}$. We now identify the time-independent constant operator $R$ with an effective Hamiltonian via  
\ba
U(\tau,0)=e^{R\tau}=e^{-\frac{i}{\hbar}\int_0^\tau H_S(t)dt}=e^{-iH_F \tau/\hbar}.
\ea
This effective Floquet Hamiltonian --- an average over a full cycle of $H_S(t)$, 
\begin{eqnarray}
 H_F&=&\frac{1}{\tau}\int^{\tau}_{0} H_S(t) dt\nonumber\\
&=&\frac{\hbar\omega_E}{2}\sigma_z^E + \frac{\hbar\omega_0}{2}\sigma_z^B + \frac{\hbar\omega_C}{2}\sigma_z^C + \frac{\hbar\omega_{EB}}{2}\sigma_z^E \sigma_z^B + \frac{\hbar\omega_{BC}}{2}\sigma_z^B \sigma_z^C + \frac{\hbar\omega_{CE}}{2}\sigma_z^C \sigma_z^E \label{H_F},
\end{eqnarray}
is defined by its quasi-energy $\hbar\omega_j$ spectrum via the relation~\cite{bhattacharjee21quantum}
\ba
 H_{F} = \sum_{j} \hbar\omega_{j} \ket{j}\bra{j}; \quad j=1,2,...,8.
 \label{hper}
\ea
Hence the part of the time-evolution operator can be decomposed as 
\begin{equation}
e^{-i H_F t/\hbar}=\sum_{j}e^{-i\omega_j t}\ket{j}\bra{j}; \quad j=1,2,...,8. 
\end{equation}
Likewise the expansion of the periodic function $P(t)$ reads~\cite{mondal2020exploring}
\ba
P(t)=\sum_{q\in\mathbb{Z}} \tilde{P}(q)e^{-iq\nu t};\quad 
\tilde{P}(q)= \frac{1}{\tau}\int_{0}^{\tau}P(t,0)e^{iq\nu t}{dt}
\ea
which results in the time-evolution operator as
\ba
U(t,0)&=&\sum_{q\in\mathbb{Z}}\sum_{j}\tilde{P}(q)e^{-i q\nu t}e^{-i\omega_j t}\ket{j}\bra{j}; \quad j=1,2,...,8. 
\label{def-u}
\ea
Thus, the system operator $\sigma^B_{\pm}(t)$ are then given by in interaction picture as~\cite{mondal2020exploring,szczygielski2013markovian}
\ba
\sigma^B_{x}(t) = U^{\dagger}(t,0) \sigma^B_{x} U(t,0)= \sum_{q \in \mathbb{Z}}\sum_{\{\Omega_B\}} (\xi(q) e^{- i\left(\Omega_B + q\nu \right)t}\sigma_{-}+\bar{\xi}(q) e^{ i\left(\Omega_B + q\nu \right)t}\sigma_{+}) ,\label{sigma-B-t}
\ea
where, 
\begin{equation}
\xi(q)=\frac{1}{\tau} \int_0^\tau \exp\left(-i\int_0^t (\omega_B(s)-\omega_0) ds\right)e^{ i q \nu t} dt.\label{xi-q}
\end{equation}
Here $\nu=2\pi/\tau$, $q$ are integers, and $\{\Omega_\alpha\}$ corresponds the set of all transition frequencies $\omega_{ij}=\omega_{i} - \omega_{j}>0$ are the possible excitation energies between the levels of $H_{F}$. It is evident due to the absence of any modulation for the emitter and collector terminals, we will get the same expressions for $\sigma^{E(C)}_{\pm}(t)$ whether we use Eq.~\eqref{H_F} or Eq.~\eqref{sys-H-s-t} in Eq.~\eqref{sigma-ec-t} for $U(t,0)$. In what folows, we  therefore write down the master equation and derive its steady state solution in terms of the time average Floquet Hamiltonian since our overall system Hamiltonian is time periodic.

With the help of Eqs.~\eqref{sigma-ec-t} and \eqref{sigma-B-t} and averaging out over 
the rapidly oscillating terms in Eq.~\eqref{reduced-master-eqn} within secular approximation, we arrive at the reduced master equation in the interaction picture~\cite{alicki2012periodically,
szczygielski2013markovian,
kosloff2013quantum,gelbwaser2015thermodynamics,
gelbwaser2015laser,gelbwaser2013minimal,
alicki2014quantum,breuer2002book,
carmichael1999book,mondal2020exploring,wijesekara2020optically} 
\ba
\dot{\rho_s}(t) &=& \mathcal{L}_{E}\left[\rho_s \right]+\tilde{\mathcal{L}}_{B}\left[\rho_s \right]+\mathcal{L}_{C}\left[\rho_s \right]; \quad  \tilde{\mathcal{L}}_{B}\left[\rho_s \right]=\sum_{q \in \mathbbm{Z}}\sum_{\{\Omega_B \}} \mathcal{L}^{B}_{q\Omega_B} \left[\rho_s\right], \quad \mathcal{L}_{E(C)}\left[\rho_s \right]=\sum_{\{\Omega_{E(C)} \}} \mathcal{L}^{E(C)}_{\Omega_{E(C)}} \left[\rho_s\right].
\label{lind}
\ea
where,
\ba
\mathcal{L}^{E(C)}_{\Omega_{E(C)}} \left[\rho_s\right] &=& G_{E(C)}(\Omega_{E(C)})\left(\sigma_{-}\rho_s(t)\sigma_{+}-\frac{1}{2}\left\{\sigma_{+}\sigma_{-},\rho_s(t)\right\}\right) +  G_{E(C)}(-\Omega_{E(C)}) \left(\sigma_{+}\rho_s(t)\sigma_{-}-\frac{1}{2}\left\{\sigma_{-}\sigma_{+},\rho_s(t)\right\}\right);\nonumber\\
\mathcal{L}^{B}_{q\Omega_B} \left[\rho_s\right] &=& P_q\Big[G_{B}(\Omega_B+q\nu)\left(\sigma_{-}\rho_s(t)\sigma_{+}-\frac{1}{2}\left\{\sigma_{+}\sigma_{-},\rho_s(t)\right\}\right) +  G_{B}(-\Omega_B-q\nu) \left(\sigma_{+}\rho_s(t)\sigma_{-}-\frac{1}{2}\left\{\sigma_{-}\sigma_{+},\rho_s(t)\right\}\right)\Big]\nonumber\\
\label{Lindblad}
\ea
and $P_q$ is the $q^{th}$-harmonic weights are calculated as $
P_q = |\xi(q)|^2 =P_{-q}$. Here we define the temperature-dependent bath response spectra  or auto-correlation function sampled as
\ba
G_B[ (\Omega_B +q \nu) ] :&=& \int_{-\infty}^{\infty} e^{ i(\Omega_B +q \nu) t} \langle \mathcal{R}_\alpha(t)\mathcal{R}_\alpha(0)\rangle dt;\nonumber\\
G_{E(C)}[ \Omega_{E(C)} ] :&=& \int_{-\infty}^{\infty} e^{ i\Omega_{E(C)} t} \langle \mathcal{R}_\alpha(t)\mathcal{R}_\alpha(0)\rangle dt,
\ea
where $\mathcal{R}_\alpha(t) = e^{i H_{R} t}\mathcal{R}_\alpha e^{-i H_{R}t}$, $\alpha\in\{E,B,C\}$. Which fulfils the detailed-balance Kubo-Martin-Schwinger (KMS) condition $G_\alpha(-\Omega) = e^{-\beta_\alpha\hbar\Omega} G_\alpha(\Omega)$~\cite{breuer2002book}. Equation~\eqref{Lindblad} stands for Eq.~\eqref{lindblad} in the main text.

\section{Derivation of steady state heat current}\label{Appendix-B}

Eight equations given by Eq.~\eqref{rho-ii-dot} are not independent since $\Tr \rho = 1$. This uniquely solves all state occupation probabilities as well as the currents. To determine the heat currents we first note that the master equation \eqref{lind} drives the system to a Gibbs-like stationary state. As stated in the previous section, since the overall system is time periodic, stationary state solutions of the master equation \eqref{Lindblad} can be expressed in the form of average Floquet Hamiltonain~\cite{gelbwaser2015thermodynamics}
\ba
\rho^{q,\Omega}_{B,ss}&=&\mathcal{Z}_B^{-1}\exp(-\frac{\Omega_B+q\nu}{\Omega_B}\beta_B H_F);\nonumber\\
\rho_{\alpha,ss}&=&\mathcal{Z}_\alpha^{-1}\exp(-\beta_\alpha H_F), \quad \alpha \in \{E,C\}
\label{rho_ss_alpha}
\ea
where $\mathcal{Z}_B= \Tr [\exp(-\frac{\Omega+q\nu}{\Omega}\beta_B H_F)]$ and $\mathcal{Z}_\alpha= \Tr [\exp(-\beta_\alpha H_F)]$. Then the thermal current or heat flow is defined by making use of the dynamical version of the second law in terms of von-Nuemann entropy~\cite{gelbwaser2015thermodynamics}. Upon taking the time derivative of the von-Neumann entropy $\mathcal{S}(\rho_s(t))= -k_B \Tr[\rho_s(t) \ln \rho_s(t)]$, one obtains
\begin{equation}\label{dev-von-neumann-entropy}
\frac{d}{dt}\mathcal{S}(\rho_s (t))=-k_B\sum_{\alpha\in \{E,C\}}\sum_{\{\Omega\}}\Tr \left[\mathcal{L}_{\Omega}^\alpha \rho_s (t) \ln \rho_s (t)\right]-k_B \sum_{q\in\mathbbm{Z}}\sum_{\{\Omega\}}\Tr \left[\mathcal{L}_{q\Omega}^B \rho_s (t) \ln \rho_s (t)\right], 
\end{equation}
where we have used Eq.~\eqref{lindblad} and $\Tr[\dot{\rho}]=0$. Now, with the help of Spohn inequality~\cite{spohn1978entropy} and dynamical version of the second law of thermodynamics~\cite{kosloff2013quantum}, one can write down the simplified version of the steady state thermal current as
\begin{eqnarray}\label{J_ss_B}
J^{ss}_B=\sum_{q \in \mathbb{Z} }\sum_{\Omega_B} \frac{\Omega_B+q\nu}{\Omega_B} \Tr \bigl[ \left(\mathcal{L}^B_{q\Omega_B} [\rho^{ss}]\right)H_F \bigr].
\end{eqnarray}
\begin{eqnarray}\label{J_ss_alpha}
J^{ss}_{\alpha}= \sum_{\Omega_\alpha} \Tr \bigl[ \mathcal{L}^{\alpha}_{\Omega_\alpha}[\rho^{ss}] H_F \bigr], \; \alpha \in \{E,C\}.
\end{eqnarray}
 Since only the base terminal is periodically modulated, only $J_B$ contains the direct signature of the modulation, while $J_{E(C)}$ do not.

Dropping the superscript and carrying out the trace over the system states, we derive the explicit expressions for the steady state heat currents using Eqs.~\eqref{rho_ss_alpha}, Eq.~\eqref{J_ss_B} and ~\eqref{J_ss_alpha}. 
\begin{eqnarray}
J_{E(C)}&=& -\hbar\sum_{\omega_{ij}} \omega_{ij}\Gamma_{ij}^{E(C)} \label{J_EC_app}\\
J_B &=& -\hbar\sum_q \sum_{\omega_{ij}} (\omega_{ij}+q\nu) \Gamma_{ij,q}^B.\label{J_B_app}
\end{eqnarray}
where the net decaying rates are as follows:
\begin{eqnarray}
\Gamma_{ij}^{E(C)} &&=[G_{E(C)} (\omega_{ij}) \rho^{ss}_{ii} - G_{E(C)} (-\omega_{ij}) \rho^{ss}_{jj}]; \nonumber \\
\Gamma_{ij,q}^{B}&&=P_q[G_{B} (\omega_{ij}+q\nu) \rho^{ss}_{ii} - G_{B} (-\omega_{ij}-q\nu) \rho^{ss}_{jj}].
\end{eqnarray}

\setcounter{equation}{0}
\renewcommand{\theequation}{C\arabic{equation}}
\section{Common Base Transistor}\label{Appendix-C}

\begin{figure}[!]
\includegraphics[width=0.95\columnwidth]{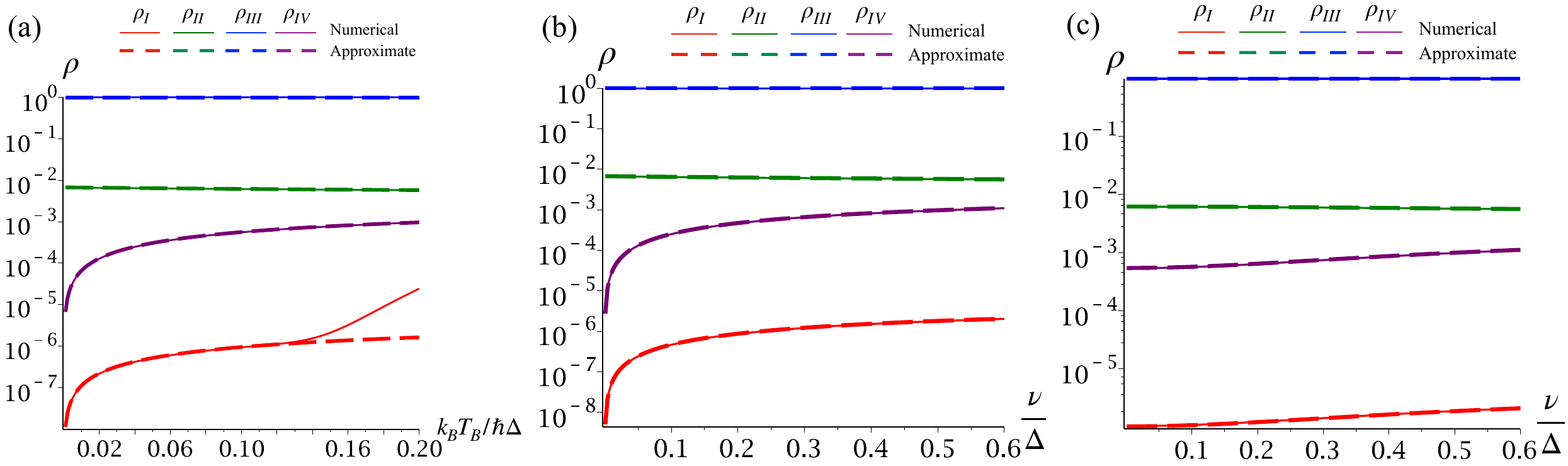}
\caption{Populations $\rho_I$, $\rho_{II}$, $\rho_{III}$ and $\rho_{IV}$ as a function of (a) $k_B T_{B}/\hbar \Delta$ for the unmodulated case, (b) as a function of $\nu/\Delta$ for $\pi$-modulation in the limit $T_{B}\rightarrow0$, and (c) as a function of $\nu/\Delta$  for $\pi$-modulation, for $k_B T_{ B}=0.118\hbar \Delta$.  Here we take $\omega_E=\omega_0=\omega_C=0$, $\omega_{CE}=0$, $\omega_{EB}=\omega_{BC}=\Delta$, $k_B T_{ E}=0.2\hbar\Delta$ and $k_B T_{C}=0.02\hbar\Delta$. As seen above, numerical results match the analytical ones as long as $T_{B}$ is not large.}
\label{FQT_fig9}
\end{figure}

\begin{figure}[!]	
\includegraphics[width=0.9\columnwidth]{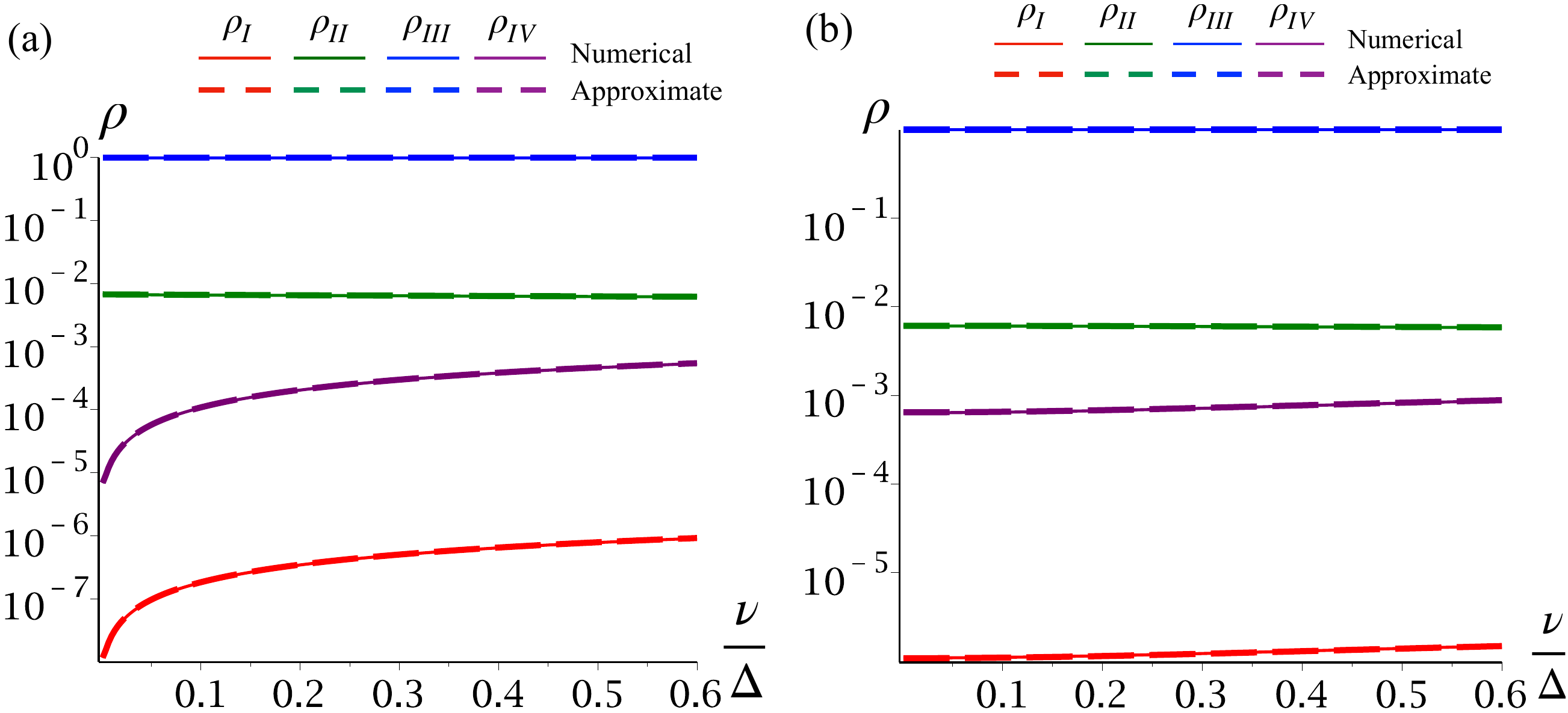}
\caption{Populations $\rho_I$, $\rho_{II}$, $\rho_{III}$ and $\rho_{IV}$ versus $\nu/\Delta$ under sinusoidal modulation with the parameters $\omega_E=\omega_0=\omega_C=0$, $\omega_{CE}=0$, $\omega_{EB}=\omega_{BC}=\Delta$, $k_B T_{ E}=0.2\hbar\Delta$ and $k_B T_{ C}=0.02\hbar\Delta$. (a) for generic case when $T_{B}\rightarrow0$ and (b) $k_B T_{ B}=0.118\hbar \Delta$. Comparison is done between populations obtained form the exact numerical calculations and the approximated analytical expressions derived here.}
\label{FQT_fig10}
\end{figure}

We now derive approximate expressions for the levels populations, thermal currents, decay rates, and analyze the conditions required for satisfaction of the Born-Markov approximation. We finally discuss the conditions needed for observing the thermal transistor effect. In order to reduce the number of the states, we set bare frequencies of all the TLSs as zero and the two non-zero couplings as symmetric 
\begin{equation}
\omega_E =0, \omega_0= 0, \omega_C=0, \omega_{EB}=\Delta, \omega_{BC}=\Delta > 0.
\end{equation}
As shown in the main text, in these conditions, the system states are degenerate two by two, viz. $|1\rangle = |8\rangle \equiv |I\rangle$, $|2\rangle = |7\rangle \equiv |II\rangle$, $|3\rangle = |6\rangle \equiv |III\rangle$ and $|4\rangle = |5\rangle  \equiv |IV\rangle$. Moreover, there are only three energy levels:  $E_1 = E_8 = \Delta$, $E_2 = E_4 = E_5 = E_7 =
0$ and $E_3 = E_6 = -\Delta$ (Fig.~\ref{FQT_fig2} of the main text). Further, we introduce new density matrix elements, now reduced to four, $\rho_{I}=\rho_{11}+\rho_{88}$, $\rho_{II}=\rho_{22}+\rho_{77}$, $\rho_{III}=\rho_{33}+\rho_{66}$, and $\rho_{IV}=\rho_{44}+\rho_{55}$. We also define the net decaying rate with respect to these new density matrix elements in the same manner. Now, we modulate the bare frequency of the base TLS in such a manner that we fix $\omega_0 =0$ i.e. the average over a time period is set to zero. Hence, on an average over time-period, the energy of the newly defined states remain same to that considered in Ref.~\cite{joulain2016quantum}. It is worthwhile to mention that the average effective Floquet Hamiltonian is essentially same as the time-independent Hamiltonian of Ref.~\cite{joulain2016quantum}. In the same spirit, for this choice of parameters, we can analyze the system to obtain the following time-evolution of the new density-matrix
\begin{eqnarray}\label{evol_rho}
&& \Dot{\rho}_I=\Gamma_{IV-I}^E + \Gamma_{III-I}^B +\Gamma_{II-I}^C \nonumber \\
&& \Dot{\rho}_{II}=\Gamma_{III-II}^E + \Gamma_{IV-II}^B +\Gamma_{I-II}^C \nonumber \\
 && \Dot{\rho}_{III}=\Gamma_{II-III}^E + \Gamma_{I-III}^B +\Gamma_{IV-III}^C \nonumber \\
 &&\Dot{\rho}_{IV}=\Gamma_{I-IV}^E + \Gamma_{II-IV}^B +\Gamma_{III-IV}^C.
\end{eqnarray}
For our problem of interest we consider $\{T_{ E},T_{ B},T_{ C}\} << \hbar\Delta/k_B$, hence $e^{\hbar\Delta/k_B T_\alpha}>>1$ which allows us to simplify the expressions of the decay rates. Finally, we obtain
\begin{eqnarray}\label{decay_rate_2}
&& \Gamma_{I-IV}^E =    \kappa \Delta \{ \rho_{I}-e^{-\hbar\Delta/k_BT_{ E}} \rho_{IV} \} \nonumber\\
&& \Gamma_{II-III}^E =    \kappa \Delta \{ \rho_{II}-e^{-\hbar\Delta/k_BT_{E}} \rho_{III} \}\nonumber \\
&&  \Gamma_{IV-III}^C  =    \kappa \Delta \{ \rho_{IV}-e^{-\hbar\Delta/k_BT_{C}} \rho_{III} \}\nonumber \\
 &&  \Gamma_{I-II}^C  =    \kappa \Delta \{ \rho_{I}-e^{-\hbar\Delta/k_BT_{C}} \rho_{II} \}.
\end{eqnarray}
Here we point out that $\Gamma_{ij}=-\Gamma_{ji}$. In the similar manner, considering up to first two harmonics we can approximate the decay rates induced by the base terminal
\begin{eqnarray} \label{decay_rate_3}
\Gamma_{I-III}^B &&=  \kappa \sum_{q=0,\pm 1}P_q (2\Delta +q\nu) \{\rho_{I}-e^{-\hbar(2\Delta+q \nu)/k_B T_{ B}}\rho_{III}\},\nonumber\\
\Gamma_{IV-II}^B 
&& = P_0 (k_B T_{B}/\hbar)[\rho_{IV}-\rho_{II}]+ P_1 \nu (2\bar{n}_B (\nu) +1)  [\rho_{IV} -  \rho_{II}].
\end{eqnarray}
Next, we would set the Eqs. \ref{evol_rho} equal to zero and solve the system of linear equations to get the steady state solutions of the density matrix elements, subject to the condition
\begin{equation}\label{trace_eq}
    Tr[\rho]=\rho_I + \rho_{II} + \rho_{III} + \rho_{IV} =1. 
\end{equation}
The set of equations can be rewritten in the form
\begin{eqnarray}\label{dynamical-eqn}
[M]\left[ {\begin{array}{c}
   \rho_I\\
   \rho_{II}\\
   \rho_{III}\\
   \rho_{IV}
  \end{array} } \right]=\left[ {\begin{array}{c}
   0\\
   0\\
   0\\
   1
  \end{array} } \right],
\end{eqnarray}
with
\begin{eqnarray}
\hspace{-1.8 cm}M=\left[ {\begin{array}{cccc}
   -2\Delta(1+P_0+2P_1) & \Delta e^{-\hbar\Delta/k_BT_{C}} & 2F(\nu)\Delta  & \Delta e^{-\hbar\Delta/k_BT_{E}}\\
   \Delta & -(\Delta+P_0 k_BT_{B}/\hbar + P_1 \nu N(\nu) + \Delta e^{-\hbar\Delta/k_BT_{C}}) & \Delta e^{-\hbar\Delta/k_BT_{ E}} & P_0k_BT_{ B}/\hbar + P_1 \nu N(\nu)\\
  2\Delta(P_0+2P_1) & \Delta & \Delta (e^{-\hbar\Delta/k_BT_{ E}} +2 F(\nu) + e^{-\hbar\Delta/k_BT_{C}}) & \Delta\\
   1 & 1 & 1 & 1
  \end{array} } \right]\nonumber\\
\end{eqnarray}
where 
\begin{eqnarray}
F(\nu)&&=  \exp(-2\hbar\Delta/k_BT_{ B})\left[P_0+P_1\left(1+\frac{\nu}{2\Delta}\right)\exp(-\hbar \nu/k_B T_{ B})+P_1\left(1-\frac{\nu}{2\Delta}\right)\exp(\hbar\nu/k_B T_{ B})\right]\nonumber\\
N(\nu)&&= \frac{1}{e^{\hbar\nu/k_B T_{ B}}-1}-\frac{1}{e^{-\hbar\nu/k_B T_{B}}-1}= 2\bar{n}^{B}_{B}+1. 
\end{eqnarray}
Under the condition of the system parameters $\{T_{ E},T_{ B},T_{C}\} << \hbar\Delta/k_B$, one can simplify the matrix $M$ following $e^{-\hbar\Delta/k_B T_\alpha}<<1$, as
\begin{eqnarray}\label{approx-M}
M \approx \left[ {\begin{array}{cccc}
   2\Delta(1+P_0+2P_1) & 0 & 0  & -\Delta e^{-\hbar\Delta/k_B T_{E}}\\
   \Delta & -\Delta-(P_0 k_B T_{B}/\hbar) - P_1 \nu N(\nu) & \Delta e^{-\hbar\Delta/k_B T_{ E}} & (P_0 k_B T_{ B}/\hbar)+P_1 \nu N(\nu) \\
  2\Delta(P_0+2P_1) & \Delta & -\Delta e^{-\hbar\Delta/k_B T_{E}}  & \Delta\\
   1 & 1 & 1 & 1
  \end{array} } \right],
\end{eqnarray}
Now from this simplified matrix Eq.~\eqref{approx-M}, the matrix determinant can also be simplified
\begin{eqnarray}\label{determinant-M}
\det[M]&\simeq& 2\Delta^2(1+P_0+2P_1)\left[\Delta+2P_0\left(\frac{k_B T_{ B}}{\hbar}\right)+2P_1\nu\left(\frac{e^{\hbar\nu/k_B T_{ B}}+1}{e^{\hbar\nu/k_B T_{ B}}-1}\right)\right](1+e^{{-\hbar\Delta}/{k_B T_{ E}}})\nonumber\\
&+&\Delta^2\left[\Delta+2P_0\left(\frac{k_B T_{ B}}{\hbar}\right)(P_0+2P_1)+2P_1\nu\left(\frac{e^{\hbar\nu/k_B T_{B}}+1}{e^{\hbar\nu/k_B T_{ B}}-1}\right)(P_0+2P_1)\right]e^{{-\hbar\Delta}/{k_B T_{ E}}}\nonumber\\
&\simeq&2\Delta^2(1+P_0+2P_1)\left[\Delta+2P_0\left(\frac{k_B T_{B}}{\hbar}\right)+2P_1\nu\left(\frac{e^{\hbar\nu/k_B T_{ B}}+1}{e^{\hbar\nu/k_B T_{B}}-1}\right)\right]\nonumber\\
&+&\Delta^2\left[\Delta(3+4P_0+8P_1)+2P_0\left(\frac{k_B T_{B}}{\hbar}\right)(2+3P_0+6P_1)+2P_1\nu\left(\frac{e^{\hbar\nu/k_B T_{B}}+1}{e^{\hbar\nu/k_B T_{ B}}-1}\right)(2+3P_0+6P_1)\right]e^{{-\hbar\Delta}/{k_B T_{ E}}}\nonumber\\
&\approx&2\Delta^2(1+P_0+2P_1)\left[\Delta+2P_0\left(\frac{k_B T_{ B}}{\hbar}\right)+2P_1\nu\left(\frac{e^{\hbar\nu/k_B T_{ B}}+1}{e^{\hbar\nu/k_B T_{ B}}-1}\right)\right]
\end{eqnarray} 
so that the r.h.s. of Eq.~\ref{determinant-M}, and therefore the populations expressions, become independent of the collector temperature $T_{\rm C}$. With Eq.~\eqref{approx-M}, evaluation of the approximated population expressions becomes straightforward:
\begin{eqnarray}
\rho_I &&\simeq \frac{1}{2(1+P_0+2P_1)} \frac{(P_0 k_B T_{ B}/\hbar) + P_1 \nu \left(\frac{e^{\hbar\nu/k_B T_{ B}}+1}{e^{\hbar\nu/k_B T_{B}}-1}\right)}{(2P_0 k_B T_{ B}/\hbar) + 2P_1 \nu \left(\frac{e^{\hbar\nu/k_B T_{ B}}+1}{e^{\hbar \nu/k_B  T_{B}}-1}\right)+\Delta}e^{-2\hbar\Delta/k_B T_{E}}\\
\rho_{II} &&\simeq  \frac{(P_0 k_B T_{B}/\hbar) + P_1 \nu \left(\frac{e^{\hbar\nu/k_B T_{ B}}+1}{e^{\hbar\nu/k_B T_{ B}}-1}\right)+\Delta}{(2P_0 k_B T_{B}/\hbar) + 2P_1 \nu \left(\frac{e^{\hbar\nu/k_B T_{B}}+1}{e^{\hbar \nu/k_B  T_{ B}}-1}\right)+\Delta}e^{-\hbar \Delta/k_{B}T_{ E}}\\
\rho_{III} &&\simeq 1- e^{-\hbar\Delta/k_B T_{E}}\\
\rho_{IV} && \simeq  \frac{(P_0 k_B T_{B}/\hbar) + P_1 \nu \left(\frac{e^{\hbar\nu/k_B T_{ B}}+1}{e^{\hbar\nu/k_B T_{ B}}-1}\right)}{(2P_0 k_B T_{ B}/\hbar) + 2P_1 \nu \left(\frac{e^{\hbar\nu/k_B T_{ B}}+1}{e^{\hbar \nu/k_B  T_{ B}}-1}\right)+\Delta}e^{-\hbar\Delta/k_BT_{ E}}.
\end{eqnarray}
In Figs.~\ref{FQT_fig9} and \ref{FQT_fig10}, we have shown the comparision between the population values calculated by their exact expressions with the analytical ones.
Similarly, we obtain the approximate decaying rates in Eqs.~\ref{decay_rate2} 
\begin{figure}[!]
\includegraphics[width=0.95\columnwidth]{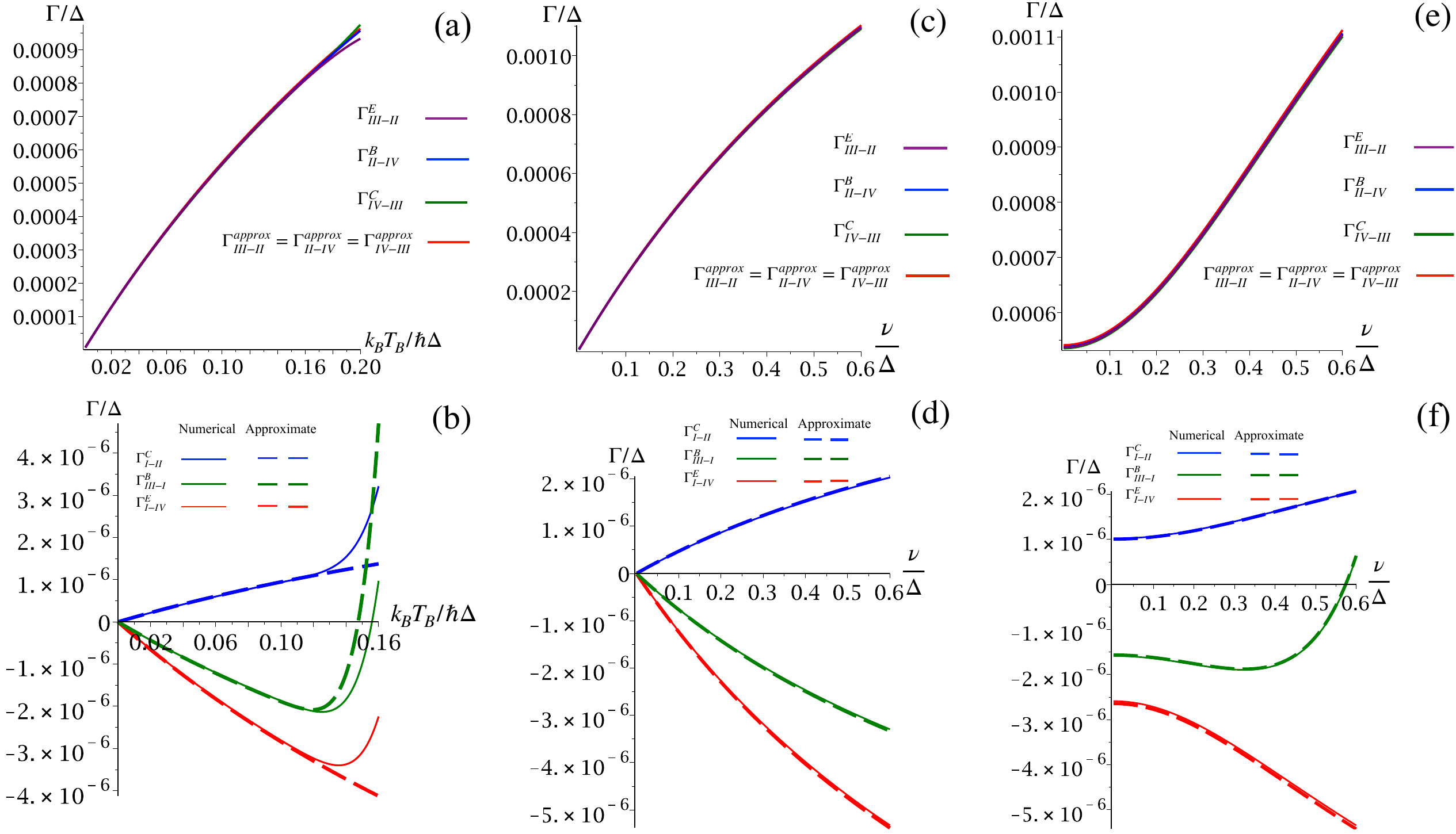}
\caption{ Decaying rates $\Gamma^{E}_{III-II}$, $\Gamma^{B}_{II-IV}$, $\Gamma^{C}_{IV-III}$ and $\Gamma^{C}_{I-II}$, $\Gamma^{B}_{III-I}$, $\Gamma^{E}_{I-IV}$ as a function of (a) and (b): $k_B T_{ B}/\hbar \Delta$ for the unmodulated case, (c) and (d): as a function of $\nu/\Delta$ in presence of $\pi$-modulation, for $T_{ B}\rightarrow0$, (e) and (f):  as a function of $\nu/\Delta$ in presence of $\pi$-modulation case for $k_B T_{ B}=0.118\hbar \Delta$ respectively. Here we take $\omega_E=\omega_0=\omega_C=0$, $\omega_{CE}=0$, $\omega_{EB}=\omega_{BC}=\Delta$, 
$k_B T_{ E}=0.2\hbar\Delta$, $k_B T_{C}=0.02\hbar\Delta$. Analytical results match the numerical ones, except for large $T_{B}$.}
\label{FQT_fig11}
\end{figure}
\begin{figure}[!]
\includegraphics[width=0.78\columnwidth, height=11.6 cm]{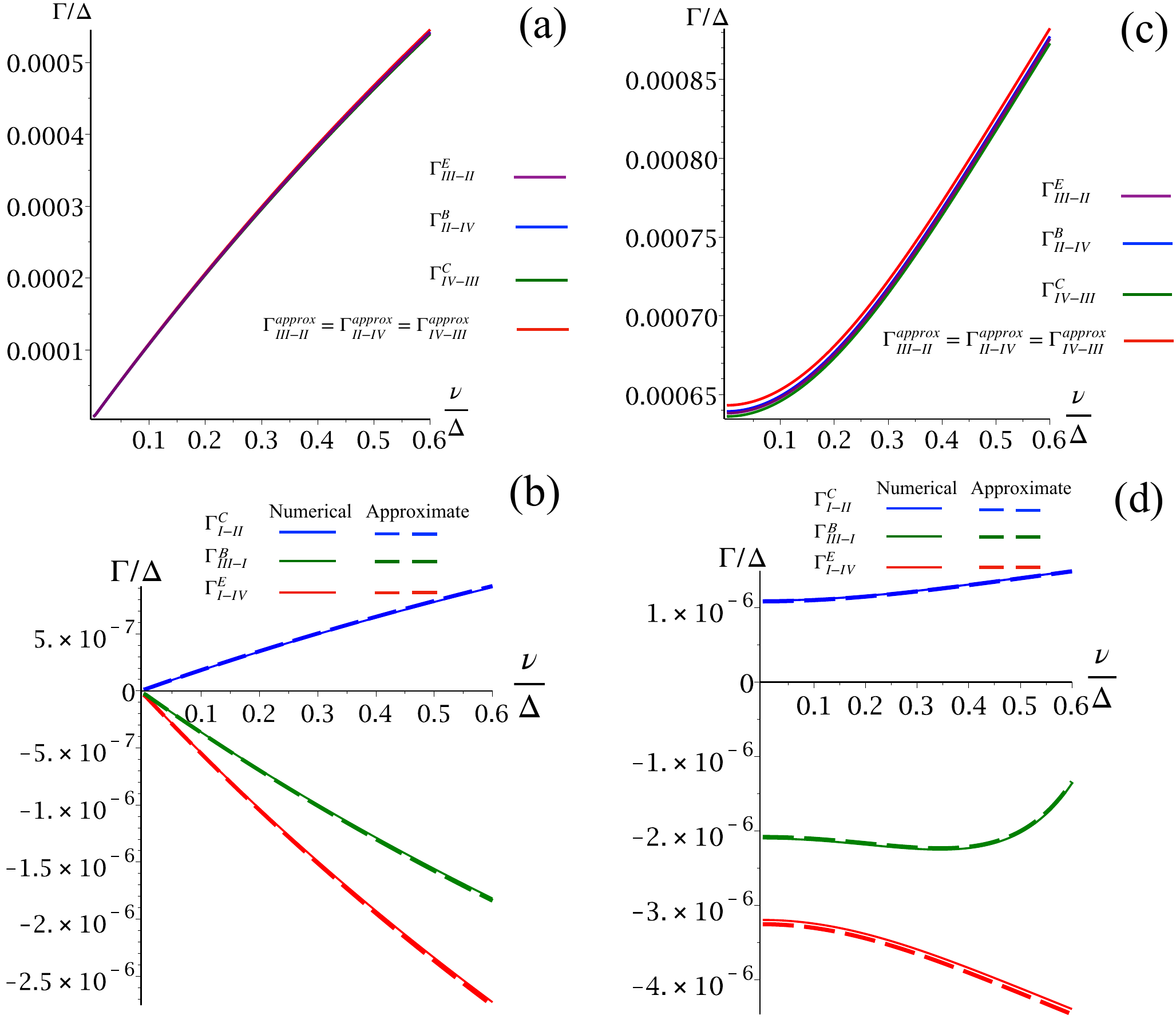}	\caption{Decaying rates $\Gamma^{E}_{III-II}$, $\Gamma^{B}_{II-IV}$, $\Gamma^{C}_{IV-III}$ and $\Gamma^{C}_{I-II}$, $\Gamma^{B}_{III-I}$, $\Gamma^{E}_{I-IV}$ as a function  $\nu/\Delta$ in presence of sinusoidal modulation, for (a) and (b) $T_{B}\rightarrow0$, and (c) and (d) for $k_B T_{B}=0.118\hbar \Delta$. Here we take $\omega_E=\omega_0=\omega_C=0$, $\omega_{CE}=0$, $\omega_{EB}=\omega_{BC}=\Delta$, $k_B T_{ E}=0.2\hbar\Delta$, $k_B T_{C}=0.02\hbar\Delta$.  Comparison is done between decay rates obtained form the exact numerical calculations and the approximated analytical expressions.}
\label{FQT_fig12}
\end{figure}

\begin{eqnarray}\label{decay_rate2}
&& \Gamma_{I-IV}^E =    -\kappa \Delta \left(\frac{1+2P_0+4P_1}{2(1+P_0+2P_1)}\right) \frac{P_0 k_B T_{B} + P_1 \hbar\nu \left(\frac{e^{\hbar\nu/{k_B T_{B}}}+1}{e^{\hbar\nu/{k_B T_{B}}}-1}\right)}{2P_0 k_B T_{B} + 2P_1 \hbar\nu \left(\frac{e^{\hbar\nu/{k_B T_{ B}}}+1}{e^{\hbar\nu/{k_B T_{ B}}}-1}\right)+\hbar\Delta}e^{-2\hbar\Delta/{k_B T_{E}}} \nonumber\\
&& \Gamma_{II-III}^E =    -\kappa \Delta \frac{P_0 k_B T_{B} + P_1 \hbar\nu \left(\frac{e^{\hbar\nu/{k_B T_{ B}}}+1}{e^{\hbar\nu/{k_B T_{ B}}}-1}\right)}{2P_0 k_B T_{B} + 2P_1 \hbar\nu \left(\frac{e^{\hbar\nu/{k_B T_{ B}}}+1}{e^{\hbar\nu/{k_B T_{B}}}-1}\right)+\hbar\Delta}e^{-\hbar\Delta/{k_B T_{ E}}}\nonumber \\
&&  \Gamma_{IV-III}^C  =    \kappa \Delta \frac{P_0 k_B T_{B} + P_1 \hbar\nu \left(\frac{e^{\hbar\nu/{k_B T_{ B}}}+1}{e^{\hbar\nu/{k_B T_{B}}}-1}\right)}{2P_0 k_B T_{B} + 2P_1 \hbar\nu \left(\frac{e^{\hbar\nu/{k_B T_{B}}}+1}{e^{\hbar\nu/{k_B T_{B}}}-1}\right)+\hbar\Delta}e^{-\hbar\Delta/{k_B T_{ E}}}\nonumber \\
&& \Gamma_{I-II}^C \simeq    \kappa \Delta \frac{1}{2(1+P_0+2P_1)} \frac{P_0 k_B T_{B} + P_1 \hbar\nu \left(\frac{e^{\hbar\nu/{k_B T_{B}}}+1}{e^{\hbar\nu/{k_B T_{ B}}}-1}\right)}{2P_0 k_B T_{B} + 2P_1 \hbar\nu \left(\frac{e^{\hbar\nu/{k_B T_{B}}}+1}{e^{\hbar\nu/{k_B T_{ B}}}-1}\right)+\hbar\Delta}e^{-2\hbar\Delta/{k_B T_{E}}} \nonumber\\
 &&  \Gamma_{I-III,q}^B =  \kappa \sum_{q=0,\pm 1}P_q (2\Delta +q\nu) \left[\frac{1}{2(1+P_0+2P_1)}\left( \frac{P_0 k_B T_{B} + P_1 \hbar\nu \left(\frac{e^{\hbar\nu/{k_B T_{ B}}}+1}{e^{\hbar\nu/{k_B T_{B}}}-1}\right)}{2P_0 k_B T_{B} + 2P_1 \hbar\nu \left(\frac{e^{\hbar\nu/{k_B T_{ B}}}+1}{e^{\hbar\nu/{k_B T_{ B}}}-1}\right)+\hbar\Delta}\right)e^{-2\hbar\Delta/{k_B T_{ E}}}-e^{-\hbar(2\Delta+q \nu)/{k_B T_{B}}}\right]\nonumber\\
&&  \Gamma_{IV-II,q}^B =  -\kappa\Delta \frac{P_0 k_B T_{B} + P_1 \hbar\nu \left(\frac{e^{\hbar\nu/{k_B T_{ B}}}+1}{e^{\hbar\nu/{k_B T_{B}}}-1}\right)}{2P_0 k_B T_{B} + 2P_1 \hbar\nu \left(\frac{e^{\hbar\nu/{k_B T_{B}}}+1}{e^{\hbar\nu/{k_B T_{B}}}-1}\right)+\hbar\Delta}e^{-\hbar\Delta/{k_B T_{E}}}.
\end{eqnarray}

An important point to note in Fig.~\ref{FQT_fig11} is that the scaled decaying rates are $\sim 10^{-6}-10^{-7}$ which are much smaller than $1$. Also from Fig.~\ref{FQT_fig12} one can see that the case of sinusoidal modulation where the decaying rates values having same order as in case of $\pi$-flip modulation. 

This implies that the system relaxation time ($1/\Gamma$) is much longer than the time scale associated with the inverse of the frequency difference ($\sim 1/\Delta$) involved in the problem, thereby justifying the Born-Markov approximation in this regime. On the other hand, larger $\Gamma$ at higher $T_{B}$ may invalidate the Born-Markov approximation (see Eq. \eqref{decay_rate_2}). Therefore we confine ourselves to low temperatures and modulate the base frequency to illustrate the transistor effect. 

Finally, we calculate the approximate form of the steady state thermal currents 
\begin{eqnarray}
&& J_E \simeq \kappa \hbar\Delta^2 \left[\frac{P_0 k_B T_{B} + P_1 \hbar\nu \left(\frac{e^{\hbar\nu/{k_B T_{ B}}}+1}{e^{\hbar\nu/{k_B T_{B}}}-1}\right)}{2P_0 k_B T_{B} + 2P_1 \hbar\nu \left(\frac{e^{\hbar\nu/{k_B T_{B}}}+1}{e^{\hbar\nu/{k_B T_{B}}}-1}\right)+\hbar\Delta}\right]e^{-\hbar\Delta/{k_B T_{E}}} \\
&& J_B = \kappa \sum_{q=0,\pm 1}P_q \hbar(2 \Delta + q \nu)^2 \left[e^{-\hbar(2\Delta+q \nu)/{k_B T_{B}}}-\frac{1}{2(1+P_0+2P_1)}\left( \frac{P_0 k_B T_{B} + P_1 \hbar\nu \left(\frac{e^{\hbar\nu/{k_B T_{B}}}+1}{e^{\hbar\nu/{k_B T_{B}}}-1}\right)}{2P_0 k_B T_{B} + 2P_1 \hbar\nu \left(\frac{e^{\hbar\nu/{k_B T_{B}}}+1}{e^{\hbar\nu/{k_B T_{ B}}}-1}\right)+\hbar\Delta}\right)e^{-2\hbar\Delta/{k_B T_{ E}}}\right] \nonumber \\
&& J_C = \simeq - \kappa \hbar\Delta^2 \left[\frac{P_0 k_B T_{B} + P_1 \hbar\nu \left(\frac{e^{\hbar\nu/{k_B T_{ B}}}+1}{e^{\hbar\nu/{k_B T_{B}}}-1}\right)}{2P_0 k_B T_{B} + 2P_1 \hbar\nu \left(\frac{e^{\hbar\nu/{k_B T_{B}}}+1}{e^{\hbar\nu/{k_B T_{B}}}-1}\right)+\hbar\Delta}\right]e^{-\hbar\Delta/{k_B T_{E}}}.
\end{eqnarray}

\end{document}